\documentclass[traditabstract]{aa}
\usepackage{natbib}
\usepackage{fancyhdr}
\usepackage{graphicx}
\usepackage{float}
\usepackage{subfigure}
\usepackage{amsmath}
\usepackage{amssymb}
\usepackage{amsfonts}
\usepackage{txfonts}
\usepackage{aas_macros}
\usepackage{dcolumn}
\usepackage{dsfont}

\usepackage{multirow}

\newcommand{\pa}[2] {\frac{\partial #1}{\partial #2}}
\renewcommand{\vec}[1]{{\boldsymbol{#1}}}

\usepackage{natbib,twoopt}
 \usepackage[breaklinks=true]{hyperref} 
 \bibpunct{(}{)}{;}{a}{}{,}    
 \newcommandtwoopt{\citeads}[3][][]{\href{http://adsabs.harvard.edu/abs/#3}%
                                        {\citealp[#1][#2]{#3}}}
 \newcommandtwoopt{\citepads}[3][][]{\href{http://adsabs.harvard.edu/abs/#3}%
                                        {\citep[#1][#2]{#3}}}
 \newcommandtwoopt{\citetads}[3][][]{\href{http://adsabs.harvard.edu/abs/#3}%
                                        {\citet[#1][#2]{#3}}}
 \newcommandtwoopt{\citeyearads}[3][][]%
   {\href{http://adsabs.harvard.edu/abs/#3}{\citeyear[#1][#2]{#3}}}

\pdfmapfile{+txfonts.map}

\begin{document}

\title{Evolution of plasma turbulence excited with particle beams}

\author{Sebastian Lange \and Felix Spanier}

\institute{Lehrstuhl f\"ur Astronomie, Universit\"at W\"urzburg, Emil-Fischer-Stra\ss e 31, D-97074 W\"urzburg \label{inst1}}
\date{}

\abstract{
Particles ejected from the Sun that stream through the surrounding plasma of the solar wind are causing instabilities. These generate wavemodes in a certain frequency range especially within shock regions, where particles are accelerated.
The aim of this paper is to investigate of amplified Alfv\'enic wavemodes in driven incompressible magnetohydrodynamic turbulence. Results of different heliospheric scenarios from isotropic and anisotropic plasmas, as well as turbulence near the \emph{critical balance} are shown. The energy transport of the amplified wavemode is governed by the mechanisms of diffusion, convection and dissipation of energy in wavenumber space. The strength of these effects varies with energy and wavenumber of the mode in question. Two-dimensional energy spectra of spherical k-space integration that permit detailed insight into the $k_\parallel/k_\perp$-development are presented.\\
The evolution of energy injected through driving shows a strong energy transfer to perpendicular wavemodes. The main process at parallel wavemodes is the dissipation of energy in wavenumber space. The generation of higher harmonics along the $k_\parallel$ axis is observed. We find evidence for a critical balance in our simulations.}

\keywords{Magnetohydrodynamics (MHD) -- Turbulence, Sun: coronal mass ejections (CMEs) -- Sun: particle emission}

\titlerunning{Turbulence excited with particle beams}
\authorrunning{S. Lange \& F. Spanier}

\maketitle

\section{Introduction}\label{sec:intro}

The Sun emits energetic particles during coronal mass ejections (CME) and solar flares. The observed energies connected to these particles reach GeVs in extreme cases. At first, solar energetic particle (SEP) events were linked exclusively to solar  flares \citepads{forbush46}, but once CMEs were first observed in the seventies, it became clear that they have a major role in the genesis of SEP events \citepads{kahler-hildner}. The  CME-driven shocks are now believed to be the major acceleration agent during the strongest SEP events \citepads[e.g.,][]{1999SSRv...90..413R}, and the most plausible acceleration mechanism in operation is diffusive shock acceleration (DSA) \citepads{bell78}. For further reading into the matter observations and models of CMEs we recommend \citetads{2006SSRv..123..251F}, and \citetads{2011LRSP....8....1C}.

In DSA, particles are accelerated through repeated crossings of the shock compression front, each crossing giving a small boost to the particle energy. The shock crossings are mediated through interactions of particles with background waves. Furthermore, the particles amplify wavemodes. This yields a coupled system of waves and particles as described, e.g., in \citetads{2007ApJ...658..622V}. Since DSA is a resonant wave-particle process, it is now interesting to see if energy transfer between wavemodes will affect different particle energies. A detailed treatement of the DSA process itself is given in \citetads{1983RPPh...46..973D}, and \citetads{Schlickeiser2002}. To enable acceleration to the GeV energies through DSA, the upstream medium needs to be very turbulent. The ambient solar wind turbulence levels are generally too low to account for this acceleration mechanism to operate beyond the MeV regime \citepads{2006GMS...165..253V}, but acceleration to the highest energies 
can still proceed through the strong amplification Alfv\'en waves in the ambient medium through streaming instabilities driven by the accelerated particles themselves. Analytical \citepads{2005ApJS..158...38L} and numerical models of
the self-consistent particle acceleration in coronal plasma have been presented \citepads{2007ApJ...658..622V,2008ApJ...686L.123N}, showing that the wave generation process is strong enough to account for the turbulence responsible for fast scattering of particles from one side of the shock to the other. The models also show that if particles are accelerated to hundreds of MeVs, the waves will grow to nonlinear amplitudes close to the shock.

It is crucial to understand the processes that govern the turbulent waves because the mechanism of DSA is strongly connected to the wave-particle interactions. Nonlinear Alfv\'en waves may interact with each other, which may lead to three important effects from the point of view of particle scattering: firstly, wave decay through three-wave interactions may limit the wave amplitudes in the shock environment; secondly, the wavenumber spectrum of the Alfv\'en waves may be altered so that the waves fall out of resonance with the particles; thirdly, the cross-helicity state of the resonant waves may also change, which affects the scattering center compression ratio of the shock and thus the accelerated particle spectrum \citepads{1998A&A...331..793V}. According to our knowledge, nonlinear wave transport models have not yet been applied to the SEP acceleration problem.

In this paper we will take the first steps towards understanding the nonlinear evolution of waves generated by energetic particle beams in the solar corona. We concentrate on a simplified model, where the beam-generated wave component is represented by a Gaussian peak in parallel wavenumber that follows the interaction of this spectral component with a quasi-isotropic background turbulence driven randomly in an incompressible magnetohydrodynamic (MHD) simulation. The turbulent plasma environment mimics the fast solar wind, where Alfv\'en waves are observed to be the dominant species \citepads{2009ApJ...706..238T,brunocarbone-livrev}, hence incompressibility can be assumed. It is especially interesting to see the energy transport in parallel and perpendicular direction. Taking into account the resonance condition, only energy transport parallel to the background magnetic field will alter the transport of particles with an energy different from the incident particle's energy. On the other hand, most turbulence theories for incompressible plasmas predict perpendicular energy transport.

\section{Numerical model}\label{sec:theory}

Despite MHD-turbulence has been studied for roughly 70 years after it was initiated by Hannes Alfv\'en, it is still a controversial topic. We  focus on incompressible turbulence for which some promising progress has been made in recent years.

One commonly observed property is the characteristic energy spectrum $E(k)$ following a power-law with a slope of $-5/3$, which is commonly referred to as the Kolmogorov--type spectrum. \citetads{kolmogorov} predicted this power-law for hydrodynamic turbulence by using dimensional analysis and scaling behaviour assuming isotropy. The basic system of turbulence evolution has also been explained by Kolmogorov. On large scales, i.e. small wavenumbers, energy is injected into the turbulent fluid. This energy decays by generating smaller structures up to the smallest scales where dissipation becomes dominant. Consequently, dissipation maintains the energy flow from small towards large $\vec k $. This is also the reason for dividing the spectrum into \emph{driving-}, \emph{inertial-}, and \emph{dissipation range}. Although Kolmogorov model of turbulence was first discussed in connection with neutral fluids, it seems to be valid in the magnetohydrodynamic case as well.

A different approach to Kolmogorov's theories by \citetads{iroshnikov} and \citetads{kraichnan} assuming a local mean magnetic field and assuming Alfv\'en wave packets led to an exponent of $-3/2$. The problem of the Iroshnikov-Kraichnan (IK) model is the assumption of isotropy because a background magnetic field will lead to a preferential direction in space caused by wave interaction resonances. The IK model implies resonant three wave interactions within a weak turbulent regime. In magnetised incompressible plasmas these interaction rates are empty (\citetads{gsweak}).

Goldreich and Sridhar (GS) describe anisotropic turbulence and distinguish between its weak \citepads{gsweak} and strong state \citepads{gsstrong}. Their assumption of strict separation between three and four wave interactions was controversially discussed and an intermediate state was introduced \citepads{gsrev}. These theories describe Alfv\'enic turbulence evolution towards the perpendicular direction. In the weak turbulence regime four wave interactions are the underlying process, referring to the GS-framework. Due to the resonance condition energy transfer to parallel wavenumbers is not possible. It is still debated wheter four wave interactions are indeed the basic mechanism in weak turbulence is questionable. In recent theories the intermediate turbulence (\citetads{gsrev}), which is based on three wave interactions, replaces the weak four wave interaction model (\citetads{lithwick2003}). However, strong turbulence is dominated by nonresonant three wave interactions, which leads to an anisotropic energy-cascade in the perpendicular direction. Parallel evolution is not caused by cascading. One of the main achievements of the Goldreich and Sridhar theory is that is explains the Kolmogorov-type energy spectrum for an anisotropic regime. This could explain the observed $-5/3$ slope in parts of the solar wind \citepads{brunocarbone-livrev} where Kolmogorov-theory is
not applicable.

The region of the heliosphere we are interested in is within the weak turbulence regime, with magnetic field fluctuations defined as
\begin{align}
 \vec{dB} \equiv \vec{B} - \vec{B_0},
\end{align}
with a mean value of $\langle \vec{dB} \rangle = 0$, which leads to  $\langle \vec{B} \rangle \sim \vec B_0$.
It is observed that the solar wind magnetic fluctuations decrease as $dB^2\propto r^{-3}$, while the background field decreases as $B^2_0\propto r^{-4}$ \citepads{1982SoPh...78..373B,brunocarbone-livrev}. Consequently, the $dB/B_0$ ratio within the heliosphere ratio is increasing with distance to the Sun \citepads{hollweg10}.

A remark on the notation, the magnetic background field $\vec{B_0}$ is defined towards the z-direction within our simulations and hence is also noted as $B_0\vec{e}_z$. The parallel direction is, therefore, defined as the z-directions and the x- and y-direction are the perpendicular directions. For symmetry reasons there will be no further distinction between the two perpendicular directions and all plots show values averaged over the azimuthal angle in cylindrical coordinates of x and y.

For small perpendicular wave numbers the transport in perpendicular direction is dominating until $k_\perp v_\perp$ is of the same order as the Alfv\'en cascading time $k_\parallel v_A$. This means that in addition to the perpendicular cascade, a cascade in the parallel direction will occur as well and energy will be transferred towards smaller parallel spatial scales. Accordingly, the parameter
\begin{align}
 \zeta \sim \frac{k_\perp v_\perp}{k_\parallel v_A}, \label{eq:zeta}
\end{align}
where $v_A$ is the Alfv\'en velocity, is of the order of unity. This state is called the \emph{critical balance} and was first introduced by \citetads{gsstrong}. In this state the linear wave period of the Alfv\'en waves are comparable to the intrisically nonlinear timescale. If $\zeta \sim 1$, the fluctuations become more correlated along the parallel direction, up to $l_\parallel \sim v_A/(k_\perp v_\perp)$ as indicated by Eq. \ref{eq:zeta}. Then the turbulence is clearly within the strong regime. This means that the fluctuations become comparable to $B_0$ and the nonlinear term is not small anymore \citepads{2008ApJ...672L..61P}.

High Reynolds numbers in combination with massive energy injection, as seen in, e.g., the solar wind, are strong indicators of a highly turbulent state. \textit{In situ} measurements of the energy spectrum \citepads{tu-marsch} agree with this fact.
To simulate conditions within the turbulent heliospheric plasma, the research group at the University of W\"urzburg has developed a simulation code, \textsc{Gismo}.

\textsc{Gismo} is an incompressible pseudospectral MHD--code that is fully parallelised and capable of efficiently using massive computing clusters.
The basis of the simulation software is to solve the following set of incompressible MHD-equations:
\begin{align}
 \pa{\vec{u}}{t} &= \vec{b} \cdot \nabla \vec{b} -\vec{u} \cdot \nabla \vec{u} -\nabla P + \nu_v \nabla^{2h} \vec{u}  \nonumber \\
 \pa{\vec{b}}{t} &= \vec{b} \cdot \nabla \vec{u} -\vec{u} \cdot \nabla \vec{b} + \eta \nabla^{2h} \vec{b} \nonumber \\
 \nabla \cdot \vec{u} &= 0 \nonumber \\
 \nabla \cdot \vec{b} &= 0, \label{eq:mhdset}
\end{align}
where $\vec{b}={\vec{B}}/{\sqrt{4\pi \varrho}}$ is the normalised magnetic field, $\vec{u}$ is the fluid velocity and $\varrho$ is the constant mass density. The diffusion coefficient related to viscous and Ohmic dissipation is denoted by $\nu_v$ and $\eta$.
A common approach in pseudospectral methods is to amplify the diffusion term by a power of $h$ - resulting in hyperdiffusivity. This artificial enhancement of the dissipation is necessary to reach a saturated state of turbulence within a reasonable timescale. It is a methodic problem, since pseudospectral approaches do not strongly suffer from dissipative numerical effects. The only intrisic energy loss of the system is caused by \emph{antialiasing}, which we discuss below.
Furthermore, the parameter $\nu$ is introduced as a global diffusivity with $\eta=\nu_v\equiv\nu$. Hence magnetic resistivity and viscous damping are not distinguishable anymore. This is the case for the magnetic Prandtl number of the order of unity, which is valid within the regime of Alfv\'en wave turbulence where an equipartion between magnetic and kinetic energy can be assumed \citepads{PhysRevE_66_046410,2008AA...490..325B}.
The pressure term $\nabla P$ fulfills the closure condition for incompressibility \citepads{marongold}
\begin{align}
 \nabla^2 P = \nabla \vec{b} : \nabla \vec{b} -\nabla \vec{u} : \nabla \vec{u}. \label{eq:pressureclosure}
\end{align}

These equations are solved in Fourier space by using pseudospectral methods that lead to the componentwise equations
\begin{align}
 \pa{\tilde u_\alpha}{t} &= -ik_\gamma\left( \delta_{\alpha \beta} - \frac{k_\alpha k_\beta}{k^2} \right) \left( \widetilde{u_\beta u_\gamma} - \widetilde{b_\beta b_\gamma} \right) - \nu k^{2h} \tilde u_\alpha, \nonumber \\
 \pa{\tilde b_\alpha}{t} &=  -ik_\beta \left( \widetilde{u_\beta b_\alpha} - \widetilde{b_\beta u_\alpha} \right)- \nu k^{2h} \tilde b_\alpha, \nonumber \\
k_\alpha \tilde u_\alpha &= 0, \nonumber \\
k_\alpha \tilde b_\alpha &= 0, \label{eq:fft-mhdset}
\end{align}
where the tilde-notation stands for quantities in Fourier space. The components of the wavevector are written as $k_\alpha$.

In the incompressible regime of a magnetised plasma the MHD-turbulence consists of only two types of waves, which propagate along the parallel direction - the so-called pseudo- and shear Alfv\'en waves. The Former are the incompressible limit of slow magnetosonic waves and
play a minor role within anisotropic turbulence (\citetads{marongold}). The pseudo Alfv\'en waves polarisation vector is in the plane
spanned by the wavevector $\vec{k}$ and $\vec{B_0}$.
The shear waves are transversal modes with  a polarisation vector perpendicular to the $\vec{k}$ - $\vec{B_0}$ plane. They are circularly polarised for parallel propagating waves.
Both species exhibit the dispersion relation $\omega^2=(v_A k_\parallel)^2$.
Note that the shear mode seems to be dominant because pseudo waves are heavily damped by the \emph{Barnes} damping process within weakly turbulent regimes \citepads{gsweak}. However the damping weakens in strong turbulence, but according to \citetads{gsstrong}, the wave generation of pseudo-Alfv\'enic wavemodes is only possible via three-wave interactions by two shear wavemodes. Barnes damping is important for high-$\beta$ plasmas. Since this is not fulfilled for the solar corona, the role of pseudo waves should not be ignored.

Because the model consists only of these two wave types, it is suitable to use a description with Alfv\'enic waves moving either forwards or backwards.
This is achieved by introducing the Els\"asser variables \citepads{elsasser}
\begin{align}
 \vec w^- &= \vec v + \vec b - v_A \vec e_\parallel \nonumber \\
 \vec w^+ &=\vec v - \vec b + v_A \vec e_\parallel,
\end{align}
and transforming the Eqs. (\ref{eq:fft-mhdset}) into a suitable form of
\begin{align}
  \left(\partial_t - v_A k_z\right) \tilde w_\alpha^- &= \frac{i}{2} \frac{k_\alpha k_\beta k_\gamma}{k^2} \left( \widetilde{w_\beta^+ w_\gamma^-} +  \widetilde{w_\beta^- w_\gamma^+}\right) \nonumber \\
		                               &-ik_\beta \widetilde{w_\alpha^- w_\beta^+} - \frac{\nu}{2} k^{2h} \tilde w_\alpha^- \nonumber \\
  \left( \partial_t + v_A k_z \right) \tilde w_\alpha^+ &= \frac{i}{2} \frac{k_\alpha k_\beta k_\gamma}{k^2} \left(\widetilde{w_\beta^+ w_\gamma^-} +  \widetilde{w_\beta^- w_\gamma^+}\right) \nonumber \\
						&-ik_\beta \widetilde{w_\alpha^+ w_\beta^-} - \frac{\nu}{2} k^{2h} \tilde w_\alpha^+. \label{eq:fft-wpm-mhdset}
\end{align}

Obviously, the nonlinearities of Eqs. (\ref{eq:fft-wpm-mhdset}) that describe the turbulent behaviour of the MHD-plasma
cannot be solved in Fourier space. Hence the main numerical load is the transfomation between real- and wavenumber space for each
iterative step. For this purpose we used the P3DFFT algorithm, which is a MPI-parallelised fast Fourier transfomation based on FFTW3 \citepads{p3dfft}.

One basic problem of spectral methods that use discrete Fourier transformation is the aliasing effect. Because of discrete sampling in the wavenumber space, high $k$-values exhibit errors that depend on the structure of the real space fields. Therefore we used zero padding, which is also referred to as Orszags $2/3$ rule, i.e. $2/3$ of the wavenumbers below the Nyquist frequency have to be truncated to achieve maximum anti-aliasing, hence reducing the Fourier space resolution to $1/3$ of the original wavenumber range (\citetads{orszag}). This process is repeated each step, immediately before calculating the nonlinearities and, accordingly, calculating the RHS of the MHD equations. Consequently, the change in the antialiasing-range of one MHD-step is physically correct, but not the long-term evolution.

\textsc{Gismo} is capable of using different foward--in--time schemes, namely Euler and Runge-Kutta second as well as fourth order. All the simulations in this paper has been peformed by using RK--4.

\begin{figure}[ht]
 \begin{center}
\includegraphics[width=0.8\columnwidth]{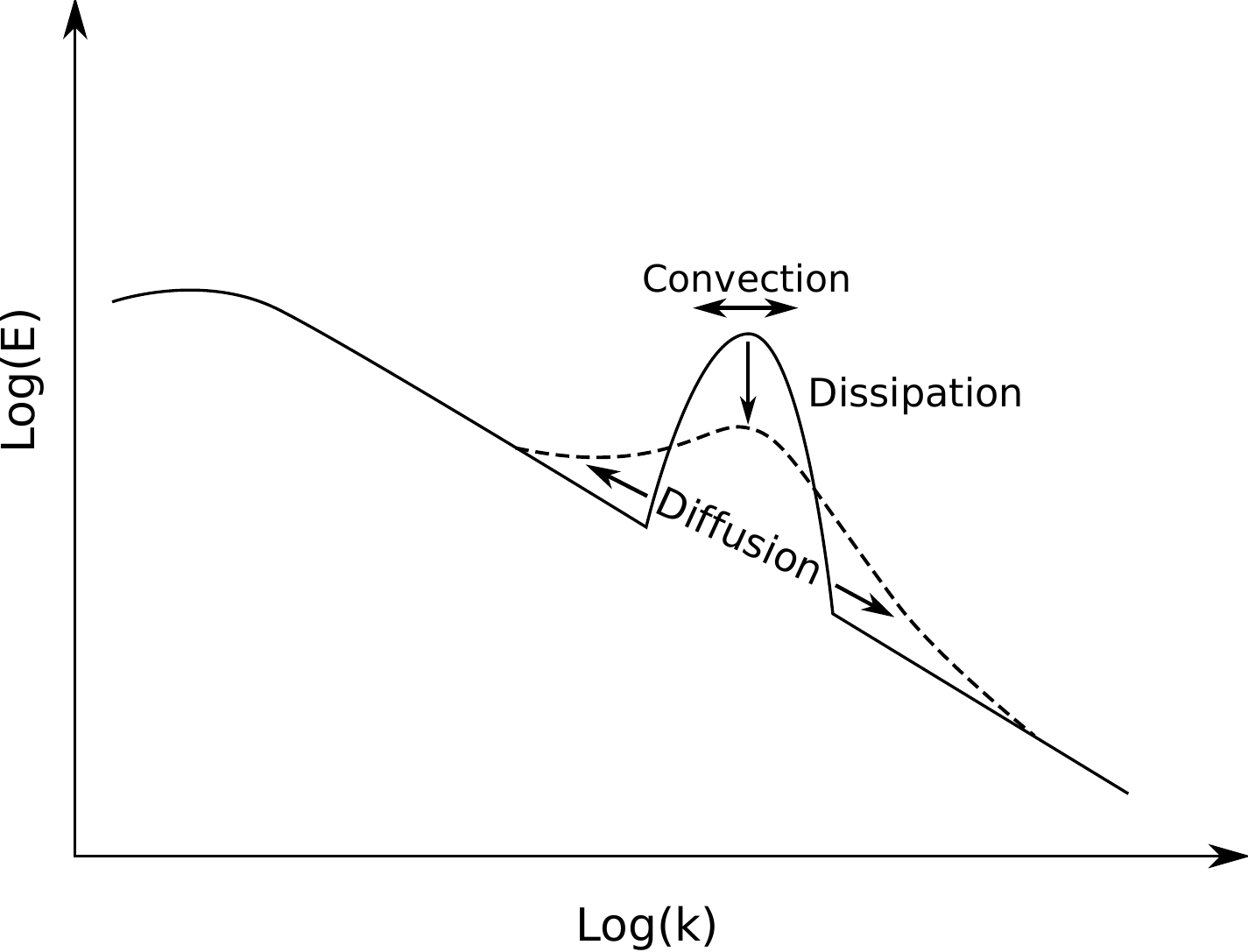} \caption{Possible mechanisms that might influence the evolution of an amplified
wavemode embedded in a turbulent plasma with a Kolmogorov-type power-law energy spectrum.} \label{fig:peak-evo-sketch}
\end{center}
\end{figure}

To mimic an SEP-event, understanding the underlying mechanisms is crucial. Before we simulated particle scattering at modified
field-fluctuations, we focussed on the evolution of amplified wavemodes within the background turbulence. Since streaming particles ejected by the Sun, e.g., protons from coronal mass ejections, will not sharply amplify a discrete mode but a broader interval, a Gaussian distributed shape was assumed. The streaming instability has the highest growth rate in the background field direction. Consequently, we assumed that only purely parallel-propagating Alfv\'en waves are modified.

The Alfv\'en wave generation mechanism by SEPs is not be investigated in detail here. The driving mechanism assumed for our simulations is the streaming instability. The estimatation of the wave growth rate is described in \citetads{rami-ongenofalfvs}. The streaming instability is caused by energetic proton scattering off the interplanetary Alfv\'en waves. During the the scattering process the particle changes its pitch angle cosine by $\Delta \mu$ while its momentum in the wave frame remains constant. Thus the particles' energy in the plasma frame is changed by $v_A p \Delta \mu$, consequently also happens to the Alfv\'en wave energy due to energy conservation. Another important instability in the solar corona is the electrostatic instability. This is caused by an electron current as well as by streaming ions. Ion acoustic waves would be generated by this process. However, for the growth rate of these modes a sufficiently high ratio $T_e/T_i \gg1$ of the electron and ion temperatures is crucial. Observations 
and simulations in the vicinity of three solar radii indicate temperature ratios of the order of unity \citepads{2007ApJ...663.1363L,2012ApJ...745....6J}. In this regime the ion acoustic waves are also efficiently suppressed by Landau damping. For further reading about the streaming instability we refer to \citetads{1993tspm.book.....G}.

These parallel peaked modes are influenced by dissipation, diffusion, and convection. An illustration of this is shown in Fig. \ref{fig:peak-evo-sketch}. Regarding the evolution within the Fourier space, dissipation will damp the wavemode, hence lower the maximum without altering the position or broadness of the peak. The dissipation of wavemodes is caused by the spatial diffusion term in Eqs. (\ref{eq:fft-wpm-mhdset}). Convection in Fourier space would shift the position of the peak. If diffusive transport in Fourier space were the dominant mechanism, it would result in a broader energy distribution. The dynamics of convection and diffusion lie within the nonlinear terms, therefore one cannot distinguish exactly between the responsible terms.

To investigate the effects of spatial diffusion on the peaks, we solved the dissipation equation in wavenumber space
\begin{align}
 \pa{e}{t}=-k^2 D e
\end{align}
and calculated its dissipation coefficient
\begin{align}
 D = \frac{\frac{e_0}{e}-1}{k^2 \Delta t}. \label{eq:diffusioncoeff}
\end{align}
When the spatial diffusion is the only dissipative effect in wavenumber space, this wavenumber dissipation coefficient would be the spatial diffusion coefficient. We emphasise that diffusion in the wavenumber space is a different process and is hence explicitly distinguished from spatial diffusion. In the discussed context above the spatial diffusion is clearly connected to the wavenumber dissipation. We concentrated our investigation on those wavemodes whose peak energy is initiated only. For other modes this approach is not feasible because the spatial diffusion is not necessarily the dominant process on an arbitrary wavemode.

\section{Simulation setup}\label{sec:simsetup}

To simulate the turbulent plasma in which the SEP-event is set, we performed the following type of magnetohydrodynamic turbulence.

\begin{table*}
\caption{ Parameter setup for the simulations.
\label{tab:simparameter}}
\begin{center}
\begin{tabular}{c c c c c c c}\hline \hline
   & $L_\text{scale}[\text{cm}]$ & $n_d[\text{cm}^{-3}]$ & $B_0[\text{G}]$ & $v_A[\text{cm } \text{s}^{-1}]$ & $\nu [\text{num}]$ & $k$-grid \\
\hline
\rule{0mm}{3mm} SI & $3.4\cdot 10^8$& $10^5$ & $0.174$ & $1.2\cdot 10^8$ & $1$ & $128^3$\\
\hline
\rule{0mm}{3mm} SII& $3.4\cdot 10^8$& $10^5$ & $0.174$ & $1.2\cdot 10^8$ & $10$ & $128^3$\\
\hline
\rule{0mm}{3mm} SIII& $3.4\cdot 10^8$& $10^5$ & $1.74$ & $1.2\cdot 10^9$ & $10$ & $128^3$\\
\hline
\rule{0mm}{3mm} SIV & $3.4\cdot 10^8$& $10^5$ & $0.00174$ & $1.2\cdot 10^{6}$ & $1$ & $128^3$\\

\hline

 \end{tabular}
\tablefoot{The wavenumber grid is defined as $k=2\pi/L_\text{scale} [\text{grid position}]$. The number density $n_d$ connects the background field $B_0$ with the Alfv\'en speed $v_A$ via $\sqrt{4 \pi m n_d}$.}
\end{center}
\end{table*}

We used an anisotropically driven turbulence with a driving range in k-space up to the first five numerical wavenumbers in perpendicular ($k'_{\perp}=2 \pi [0\cdots4 ] $) and 15 in parallel direction ($k'_{\parallel}=2 \pi [0\cdots14 ] $). A remark to the notation, the wavenumber is in general defined as $k = (2 \pi n)/L$ where $n$ stands for the numerical grid position. For simplicity we used the normalised wavenumbers $k' = (2 \pi \cdot n)$ throughout. The anisotropy was chosen for two reasons. First, to mimic the preferential direction of the solar wind, where particles that radially stream away from the Sun form the Parker spiral. Consequently, these particles can deposit their energy in a parallel direction on different scales. This is mainly valid in the vicinity of the Sun, in which we are interested. Second, a slab--component of SW turbulence is observed also at small scales in parallel direction. To ensure turbulence evolution up to high parallel 
wavenumbers, the driving range was extended along the parallel axis. This is necessary because the parallel evolution is much weaker than the perpendicular. Even though this is primarily a technical aspect to ensure the extent of the spectrum to higher $k_\parallel$, it is still in line with observations. An isotropic driver would not yield sufficiently turbulent modes at high $k_\parallel$.
The turbulence driving is performed by allocating an amplitude with a phase to the Els\"asser--fields within the Fourier space. The amplitude follows a power-law of $|\vec{k}|^{-2.5}$ and is initialised using a random normal distribution. The phase was randomly chosen between zero and $2\pi$. These settings are divergence-free and hermitian symmetric. After this initialisation the values were scaled to the desired scenario, which in our case is a $dB/B_0$ ratio of roughly $10^{-2}$. Note that both species, pseudo- and shear Alfv\'en waves, are excited by this type of turbulence driving, but as presented by \citetads{marongold} the pseudo-waves evolution is strongly suppressed.
In this initial range energy is injected at discrete times, which leads to a saturated turbulence - an
equilibrium between dissipation and injection. The spatial resolution is $256^3$ gridpoints, resulting in $128^3$ points in k-space of which $|\vec{k'}|=2 \pi \cdot 42$ wavemodes are active modes that remain unaffected by (anti)aliasing. The hyperdiffusivity coefficient was set to $h=2$.

The simulations of the turbulent background plasma were performed assuming an outer scale of $L_\text{scale}=3.4\cdot 10^8 \text{cm}$. This value was estimated using the growth rate from \citetads{rami-ongenofalfvs}
\begin{align}
 \Gamma (k) = \frac{\pi \omega_{cp}}{2 n_p v_A} \int \text{d}^3 p \text{ } v\; \mu |k|  \text{ } \delta(|k|-\frac{\omega_{cp}}{\gamma v_p}) \, f_p,
\label{eq:growthraterami}
\end{align}
with the proton cyclotron frequency $\omega_{cp}$, the proton speed $v_p$, the Lorentz factor $\gamma$, the proton number density $n_p$, $\mu$ the pitch angle cosine, and $f_p$ the proton distribution function. Here the resonance condition for the $n$th order of interaction
\begin{align}
 k_{\parallel} v_{\parallel} - \omega - n \Omega = 0, \quad n \in \mathds{Z}
\label{eq:wave-particle-res}
\end{align}
(cf. \citealt{schlickeiser89}) was used, where $\omega$ is the wave frequency, and $ k_{\parallel}$ its parallel wavenumber. $\Omega$ is the particle's gyrofrequency and $v_{\parallel}$ its parallel velocity component. Note that Eq. \ref{eq:growthraterami} is only valid for purely parallel waves and $n=\pm 1$. Orders of $|n|>1$ can only be generated by oblique waves. The perpendicular components of the wave would then modify the scattering process by nonvanishing Bessel functions. This is discussed in detail by \citetads{Schlickeiser2002}.
We used peaks at $k=2\pi\cdot 8$ and $k=2\pi\cdot24$, which represent proton energies of $E\approx 64 \text{ MeV}$ and $E\approx 7 \text{ MeV}$ respectively. Using the resonance condition, this leads to a length scale of
\begin{align}
 L_\text{scale}= \frac{2 \pi n}{e B_0} \gamma m_p c v \approx 10^8 \text{ cm}.
\end{align}

The Alfv\'en speed was assumed to be $v_A = 1.2 \cdot 10^8 \text{ cm}\text{ s}^{-1}$, which leads along with a particle number density of $10^5 \text{ cm}^{-3}$ to a background magnetic field $B_0 = 0.174 \text{ G}$. These values resemble the solar wind environment at three solar radii (\citetads{ramigramm}), where particle acceleration by CME--driven shocks is strongest.

The discretisation of the timestep is stable for values up to $\Delta t' = 1 \cdot 10^{-11}$ in numerical units or $\Delta t' = 3.4 \cdot 10^{-3}\text{ s}$ for the background turbulence.

If the background plasma simulation reaches the saturated state, a Gaussian--distributed energy peak with purely parallel $\vec{k}=k \vec{e_\parallel}$ is injected. We chose two different positions of the peak in wavenumber space. To investigate the physics of an SEP--event, a wavenumber of $k_\parallel=1.5\cdot10^{-7} \text{ cm}^{-1}$ was used. This corresponds to a numerical wavenumber of 8, which is still within the driving range of the turbulence. The injection at smaller scales was represented by a peak at $k_\parallel=4.4\cdot10^{-7} \text{ cm}^{-1}$. This value lies at the numerical position 24, which is roughly between the maximum driven wavemode and the anti-aliasing truncation edge (which was at 43).
We injected the SEP-energy gradually over a certain time interval to develop a realistic scenario. Multiple situations were explored, by using simulations with peaks at either position, with either large (growth rate $\Gamma_1$) or small (growth rate $\Gamma_2$) total amplitude of the Gaussian at the final driving step. Because the velocities increase near the peaks, the discretisation of the timestep had to be decreased to values of  $\Delta t' = 5 \cdot 10^{-12}$ in numerical units or $\Delta t' = 1.7 \cdot 10^{-3}\text{ s}$ to sustain stability.

To allow even more diverse case studies, four different initial conditions were used, as described in table \ref{tab:simparameter}. In each setup a complete evolution of the background turbulence was simulated. The first setup SI uses the standard parameters as described above. The main parameters of interest are the resistivity of the plasma and the background magnetic field. The results in changing these will reveal the effects on the mechanisms described in Fig. \ref{fig:peak-evo-sketch}. An increased resistivity compared to simtype SI is approached in SII. A higher value for $\nu$ is expected to make a difference in the spatial diffusion behaviour and would make wavenumber dissipation more dominant to the other transport. As indicated in Fig. \ref{fig:peak-evo-sketch}, this would lead to a significant damping of the peak. The dissipation range of the background turbulence will likely be increased by this parameter as well.
The third setup SIII has a magnetic field increased by a factor of 10. This is to examine the influence of a more anisotropic turbulence because the perpendicular evolution should be much stronger according to \citetads{gsstrong}. In general, these values may only be achieved in magnetic clouds, but it gives valuable information on the mechanisms of turbulent transport. The high resistivity is necessary because of stability problems with the accompanying high Alfv\'en speeds.
The last variation of the scenario, which was used in SI, is a strongly decreased magnetic background field $B_0$, see simtype SIV. The aim of this artificial scenario is to investigate strong turbulence at $\zeta \approx 1$.

\section{Results}\label{sec:results}

\begin{figure}[ht]
 \begin{center}
\includegraphics[width=\columnwidth]{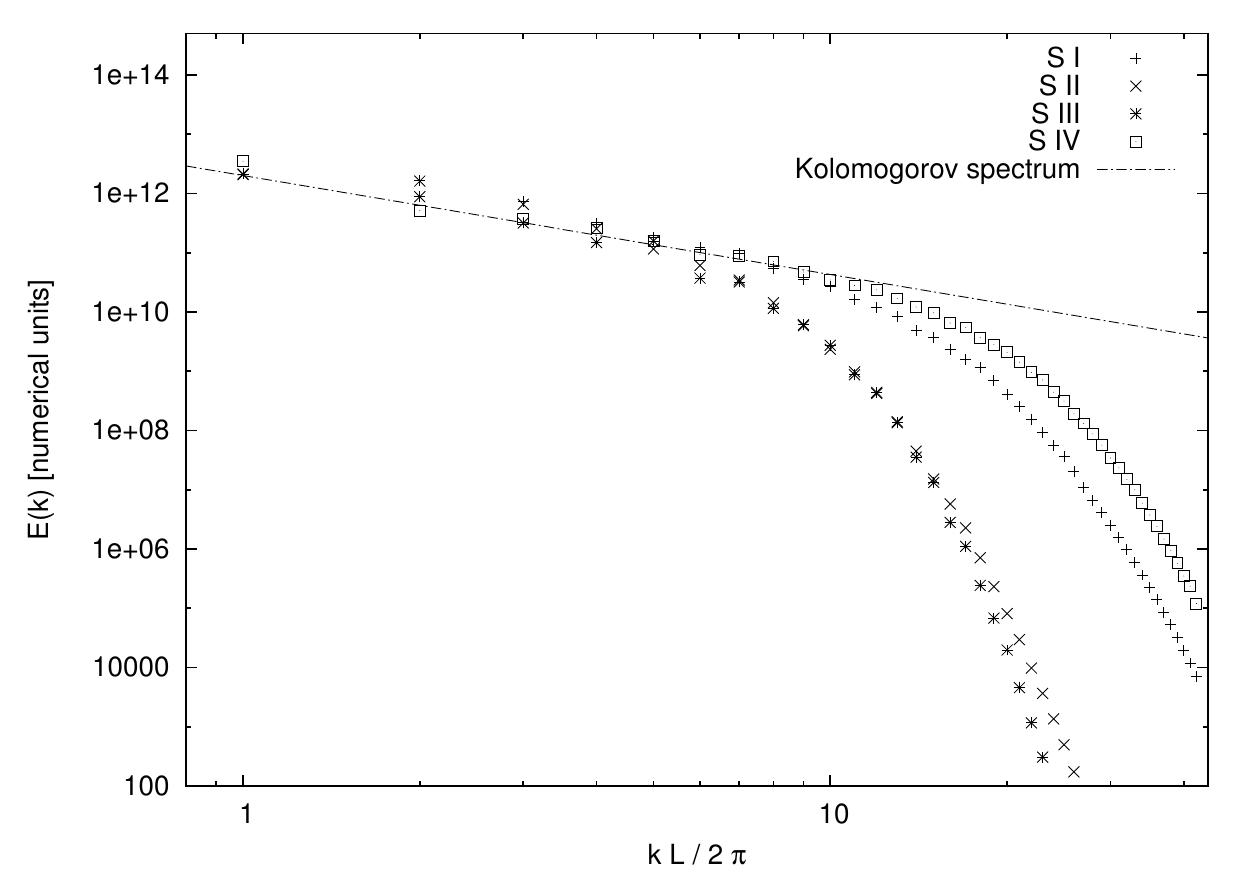} \caption{Magnetic energy spectrum of the simulated background turbulence setups. The plot was made by total integration within the Fourier space.} \label{fig:backgroundturb}
\end{center}
\end{figure}

The evolution of the anisotropic background turbulence was simulated up to 30 Alfv\'en wave crossing times, which corresponds to a simulation of 85 s in physical time.
At this point, the turbulence has reached a saturated state and a Kolmogorov-type power-law has evolved over a wide range of wavenumbers (see Fig. \ref{fig:backgroundturb}). According to \citetads{gsweak} and \citetads{gsstrong}, the 5/3-spectrum is dominant for perpendicular $k_\perp$, whereas the parallel evolution is significantly slower. Our simulations confirm this behaviour very clearly. Note that the total spectra in Fig. \ref{fig:backgroundturb} deviate from the Kolmogorov shape, especially at high wavenumbers, because the shape of the parallel spectrum is not 5/3. As expected, the spectra are very sensible to $\nu$. A factor of ten increases the dissipation range drastically. This is also due to the hyperdiffusivity we used where higher wavenumbers are damped by higher power of k (dissipation term $\propto k^4$) (see Sect. \ref{sec:theory}).
The $dB/B_0$ ratio of the developed turbulence is about $10^{-2}$. The magnetic field fluctuations are defined in Fourier space by
\begin{align}
 (dB)^2 = \int \limits_{|\vec{k}|>0} \text{d}^3 k \, \frac{1}{4}(\tilde w^-(\vec{k}) - \tilde w^+(\vec{k}))^2.
\end{align}
The influence of the turbulence strength on the energy transport is an important aspect for the peak simulations. We show this below.

\begin{table}
\caption{ Evolution timesteps of the peaks.
\label{tab:peaktimes}}
\begin{center}
\begin{tabular}{c c c c }\hline \hline
  $t[\text{s}]$ & $k'_\text{Peak}= 2\pi \cdot 8$ & $k'_\text{Peak}= 2\pi \cdot 24$ & \\
\hline
$t_1 \equiv t_\text{start}$ & 0 & 0 & \multirow{3}{*}{SI} \\
$t_2 \equiv t_\text{mid}$ & $51 \, (25.5)$ &  $8.5$ & \\
$t_3 \equiv t_\text{end}$ & $102 \, (51)$ & $17$ & \\
\hline
$t_1 \equiv t_\text{start}$ & 0 & 0 & \multirow{3}{*}{SII} \\
$t_2 \equiv t_\text{mid}$ & $20.4$ &  $1.28$ & \\
$t_3 \equiv t_\text{end}$ & $40.8$ & $2.55$ & \\
\hline
$t_1 \equiv t_\text{start}$ & 0 & 0 & \multirow{3}{*}{SIII} \\
$t_2 \equiv t_\text{mid}$ & $22.53 \, (21.42)$ &  $2.04$ & \\
$t_3 \equiv t_\text{end}$ & $45.05 \, (42.84)$& $4.08$ & \\
\hline
$t_1 \equiv t_\text{start}$ & 0 & 0 & \multirow{3}{*}{SIV} \\
$t_2 \equiv t_\text{mid}$ & $54 \, (28.9)$ &  $12.75$ & \\
$t_3 \equiv t_\text{end}$ & $108 \, (57.8)  $& $25.5$ & \\
\hline

 \end{tabular}
\tablefoot{The labels \emph{start}, \emph{mid} and \emph{end} stand for the times of the decay until the final dissipation of the peak mode. The time $t_\text{mid}$ is defined as the half-time of the decay cycle. Note that the peak at smaller wavenumber $k'_\parallel=2\pi \cdot 8$ remains  visible significantly longer than the other peak. The values denoted in brackets are the middle and end of the simulations that were not performed until the final dissipation of the peak due to the long computational times. In these cases we estimated of the total decay by an exponential fit of the decay curve.}
\end{center}
\end{table}

\begin{figure}[ht]
 \begin{center}
\includegraphics[width=\columnwidth]{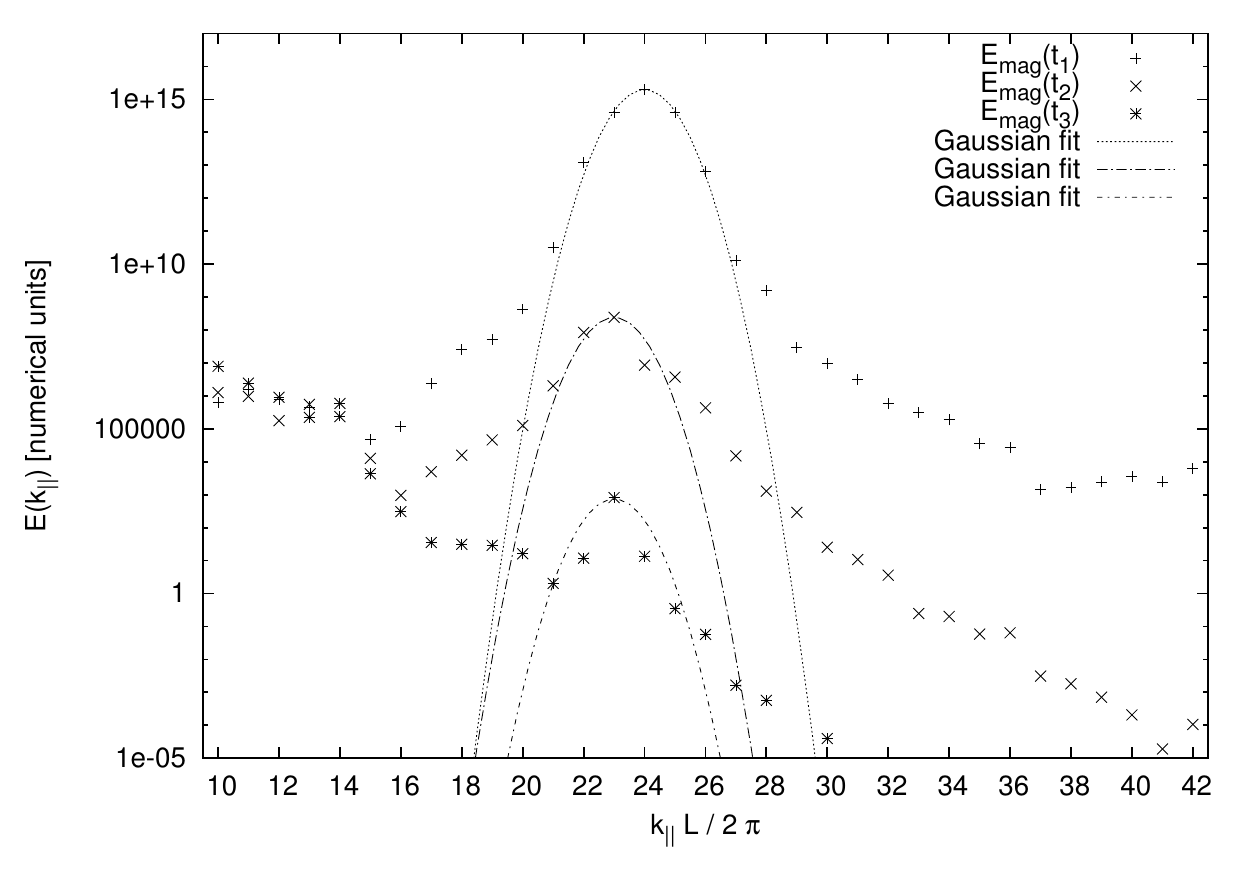} \caption{Simulation setup SI: Time evolution of a Gaussian--distributed amplification
 at numerical wavenumber 24 ($\hat= 4.4\cdot 10^{-7} \text{ cm}^{-1}$) at the lower growth rate $\Gamma_1$. The spectrum is a one--dimensional cut along the parallel wavenumber axis where the peak is located. The peak is clearly shifted towards smaller $k_\parallel$.} \label{fig:peak24smallgauss}
\end{center}
\end{figure}

\begin{figure}[ht]
 \begin{center}
\includegraphics[width=\columnwidth]{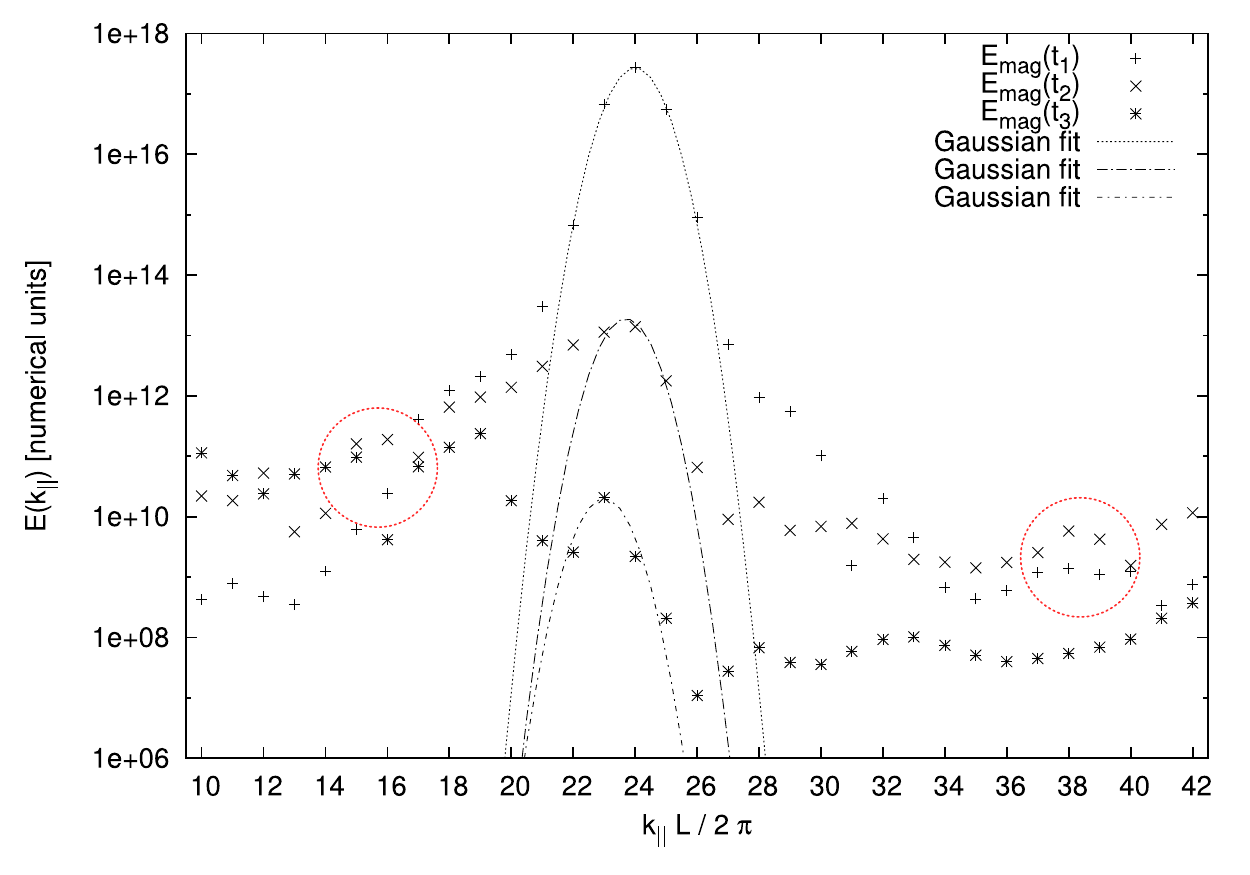} \caption{Simulation setup SI: One--dimensional energyspectrum $E(k_\parallel)$ of the peak at numerical wavenumber 24 ($\hat= 4.4\cdot 10^{-7} \text{ cm}^{-1}$) with higher growth rate $\Gamma_2$. The influence of the diffusion process is significant because the peak is broadening during time evolution. Adjoining maxima develop, e.g. at $k'_\parallel = 2 \pi \cdot 16 $ and 38, highlighted by red circles.} \label{fig:peak24biggauss}
\end{center}
\end{figure}

Once the background plasma was simulated, a peak is driven over a time period of ca. 1.7 s.
An exemplary time evolution of the peak at normalised wavenumber 24 is shown in Figs. \ref{fig:peak24smallgauss} and \ref{fig:peak24biggauss}.
This is a one-dimensional spectrum of the magnetic field energy in numerical units that shows a cut along the parallel axis.
The starting time $t_\text{start}$ corresponds to the end of the driving interval, hence the maximum amplification of the wavemode.
The timesteps are shown in table \ref{tab:peaktimes}. For subsequent use the times $t_\text{mid}$ and $t_\text{end}$ are introduced. Note that the decay time interval of the excited mode at $k'_\parallel=2\pi \cdot 8$ is significantly longer compared to the $k'_\parallel=2\pi \cdot 24$ peak. This is again because of the hyperdiffusivity, which damps higher modes more strongly because the dissipation term is $\propto k^4$. The time $t_\text{end}$ is defined at the state where the peak has lost nearly its complete energy within the parallel direction. We took this as the final point of the energy diffusion or dissipation.

\begin{figure}[ht]
 \begin{center}
\includegraphics[width=\columnwidth]{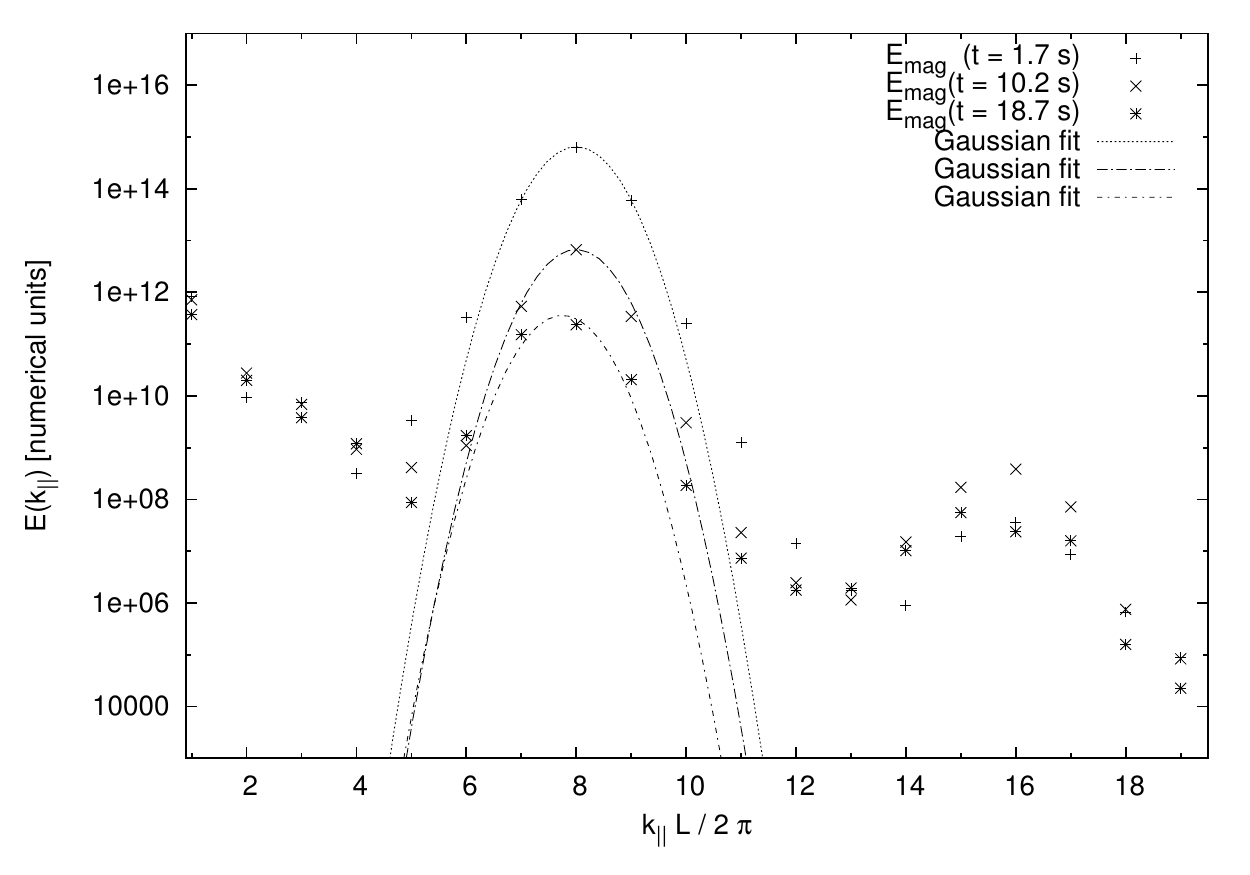} \caption{Simulation setup SI: Time evolution of a Gaussian--distributed amplification at numerical wavenumber 8 ($\hat= 1.5\cdot 10^{-7} \text{ cm}^{-1}$) at the lower growth rate $\Gamma_1$. The spectrum is a one--dimensional cut along the parallel wavenumber axis where the peak is located. Only a slight shift towards small $k_\parallel$ is observed. The broadening is clearly visible, especially on the flanks of the Gaussian curve.} \label{fig:peak8smallgauss}
\end{center}
\end{figure}

Both peaks at $k'_\parallel=2\pi \cdot 24$ show broadening of its shape and shifting from the initial value to $k'_\parallel=2\pi \cdot 23$ within 17 s, while peaks at $k'_\parallel=2\pi \cdot 8$ are only slightly shifted to $k'_\parallel=2\pi \cdot 7.7$ s within 17 s, but the broadening is clearly visible, as shown in Fig. \ref{fig:peak8smallgauss}.

The next step is to investigate the development of the amplified wavemodes. If the evolution of the peak were solely governed by spatial diffusion, the dissipation coefficient $D$ of Eq. (\ref{eq:diffusioncoeff}) would stay constant in time. Therefore the change in peak energy was measured.
Two intervals were used for the calculation of $D$, denoted as time-interval $\tau_1$ and $\tau_2$. These intervals are different for both peak postions because of the faster decay at high wavenumbers. At $k'_\parallel = 2 \pi \cdot 8$ the time intervals are $\tau_1 = 8.5$ s and $\tau_2 = 17$ s, whereas at $k'_\parallel = 2 \pi \cdot 24 $ the intervals $\tau_1 = 1.7$ s and $\tau_2 = 3.4$ s are used.
The results of the diffusion coeffients are given in table \ref{tab:diffcoeffs}.

\vspace{0.5cm}
\begin{table}
\caption{Dissipation coefficients according to Eq. (\ref{eq:diffusioncoeff}).
\label{tab:diffcoeffs}}
\begin{center}
 \begin{tabular}{ c c c c c }\hline \hline
  \rule[-0.15cm]{0pt}{0.5cm} &\multicolumn{2}{c }{$k'_\text{Peak}= 2\pi \cdot 8$} & \multicolumn{2}{c}{$k'_\text{Peak}=2\pi \cdot24$ } \\

&$\Gamma_1$ & $\Gamma_2$ & $\Gamma_1$ & $\Gamma_2$\\
\hline
\multicolumn{5}{c}{\textsc{Simtype SI}\rule[-0.2cm]{0pt}{0.5cm}} \\

$\tau_1$\rule[-0.15cm]{0pt}{0.5cm}& \rule{0.5mm}{0mm} $5.09\cdot10^{14}$ \rule{0.5mm}{0mm}& \rule{0.5mm}{0mm}$1.76\cdot10^{16}$ \rule{0.5mm}{0mm}&\rule{0.5mm}{0mm} $4.12\cdot10^{15}$\rule{0.5mm}{0mm} & \rule{0.5mm}{0mm} $7.92\cdot10^{15}$ \rule{0.1mm}{0mm}\\

$\tau_2$\rule[-0.15cm]{0pt}{0.5cm}& $7.23\cdot10^{15}$ & $8.79\cdot10^{16}$ & $1.55\cdot10^{16}$ & $2.92\cdot10^{16}$\\
\hline
\multicolumn{5}{c}{\textsc{Simtype SII}\rule[-0.2cm]{0pt}{0.6cm}} \\

$\tau_1$\rule[-0.15cm]{0pt}{0.5cm}& $1.41\cdot10^{15}$ & $8.33\cdot10^{15}$ & $9.79\cdot10^{24}$ & $1.23\cdot10^{25}$\\

$\tau_2$\rule[-0.15cm]{0pt}{0.5cm}& $3.28\cdot10^{17}$ & $7.60\cdot10^{15}$ & $4.85\cdot10^{30}$ & $4.91\cdot10^{32}$\\
\hline
\multicolumn{5}{c}{\textsc{Simtype SIV}\rule[-0.2cm]{0pt}{0.6cm}} \\

$\tau_1$\rule[-0.15cm]{0pt}{0.5cm}& $1.25\cdot10^{14}$ & $6.74\cdot10^{13}$ & $5.84\cdot10^{14}$ & \dots \\

$\tau_2$\rule[-0.15cm]{0pt}{0.5cm}& $4.43\cdot10^{14}$ & $8.00\cdot10^{13}$ & $1.47\cdot10^{16}$ & \dots \\

\hline

 \end{tabular}
\tablefoot{The coefficients are calculated for different simulation types (SI, SII, SIV) at different time intervals ($\tau_1, \tau_2$), for different growth rates ($\Gamma_1, \Gamma_2$), and for the two driving wave numbers ($k=2\pi\cdot 8, 2\pi\cdot 24$). Not calculated are the coefficients for SIII because two parameters were changed in this setup. Therefore it is not directly comparable to the other simulations. The diffusion coefficients are given in $\text{cm}^{2}\text{s}^{-1}$. }
\end{center}
\end{table}
\vspace{0.5cm}

\begin{table}

\caption{Energy deposited at the driven peak wave number.
\label{tab:peakamplitudes}}
\begin{center}
\begin{tabular}{c c c c}\hline \hline
\multicolumn{2}{c}{  \rule[-0.15cm]{0pt}{0.5cm}$k'_\text{Peak}= 2\pi \cdot 8$} & \multicolumn{2}{c}{$k'_\text{Peak}=2\pi \cdot24$ } \\

$\Gamma_1$ & $\Gamma_2$ & $\Gamma_1$ & $\Gamma_2$\\
\hline
\multicolumn{4}{c}{\textsc{Simtype SI}\rule[-0.2cm]{0pt}{0.6cm}} \\

\rule[-0.15cm]{0pt}{0.5cm} \rule{1.mm}{0mm} $1.90\cdot10^{7}$ \rule{1.mm}{0mm}& \rule{1.mm}{0mm}$1.95\cdot10^{9}$ \rule{1.mm}{0mm}&\rule{1.mm}{0mm} $1.66\cdot10^{18}$\rule{1.mm}{0mm} & \rule{1.mm}{0mm} $2.29\cdot10^{20}$ \rule{1.mm}{0mm}\\
\hline
\multicolumn{4}{c}{\textsc{Simtype SII}\rule[-0.2cm]{0pt}{0.6cm}} \\

\rule[-0.15cm]{0pt}{0.5cm} \rule{1.mm}{0mm}$9.17\cdot10^{7}$\rule{1.mm}{0mm} & \rule{1.mm}{0mm}$9.22\cdot10^{9}$ \rule{1.mm}{0mm}& \rule{1.mm}{0mm}$2.10\cdot10^{18}$ \rule{1.mm}{0mm}& \rule{1.mm}{0mm}$2.10\cdot10^{20}$\rule{1.mm}{0mm}\\
\hline
\multicolumn{4}{c}{\textsc{Simtype SIII}\rule[-0.2cm]{0pt}{0.6cm}} \\

 \rule[-0.15cm]{0pt}{0.5cm}$4.41\cdot10^{7}$ & $4.39\cdot10^{9}$ & $4.98\cdot10^{19}$ & $4.98\cdot10^{21}$\\
\hline
\multicolumn{4}{c}{\textsc{Simtype SIV}\rule[-0.2cm]{0pt}{0.6cm}} \\

\rule[-0.15cm]{0pt}{0.5cm} $1.45\cdot10^{6}$ & $1.54\cdot10^{8}$ & $5.17\cdot10^{9}$ & \dots \\

\hline

 \end{tabular}
\tablefoot{The energy is given as the ratio of $E(k_\text{Peak},t_\text{max})/E(k_\text{Peak},t_\text{start})$.}
\end{center}
\end{table}

To relate the growth rates $\Gamma_{1/2}$ with the total resulting amplitude of the Gaussians of the simulations SI-SIII, the energy was measured and compared to the background at timestep $t_\text{start}$ each. The results are presented in table \ref{tab:peakamplitudes}.

\begin{figure*}[ht]
 \begin{center}
\includegraphics[width=\textwidth]{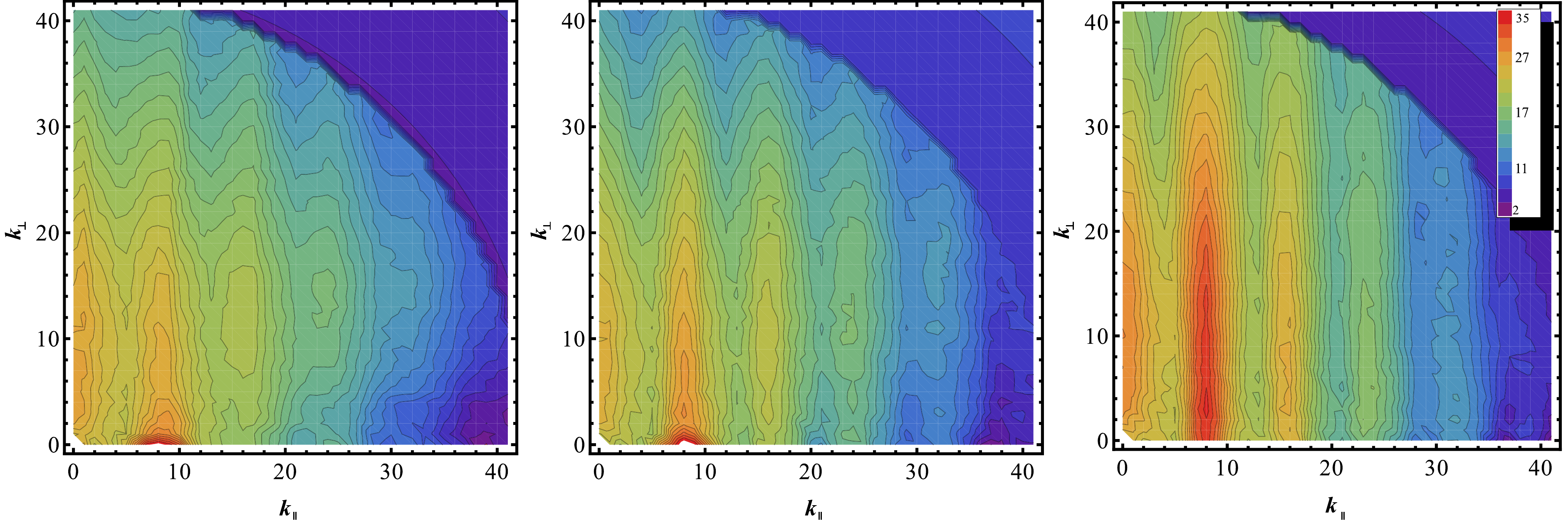} \caption{Two-dimensional magnetic energy spectra of a peak at normalised wavenumber 8.
Red regions are at higher energies compared to the blue ones. The parallel and perpendicular wavenumbers are given as absolute values.
The time development is shown for mid-drive ($\Delta t \approx 0.85 \text{ s}$), max-peak ($\Delta t \approx 1.7 \text{ s}$) and the decay 17 s after the driving. The colours of the contours were normalised for comparison between the three plots. The colours indicate the logarithm of the total spectral energy. The simulation setup SI was used.}
\label{fig:sphereplot-timeevo}
\end{center}
\end{figure*}

To investigate the direction of the peak evolution in the parallel and perpendicular directions, two--dimensional contour plots were produced. Fig. \ref{fig:sphereplot-timeevo} shows the time evolution of a peak at normalised $k'_\parallel=2 \pi \cdot  8$. The two--dimensional spectrum is a contour plot of the power spectral density of the magnetic field that was calculated by cylindrical integration in $k$-space. The single contours are scaled logarithmically.

\begin{figure}[ht]
 \begin{center}
\includegraphics[width=\columnwidth]{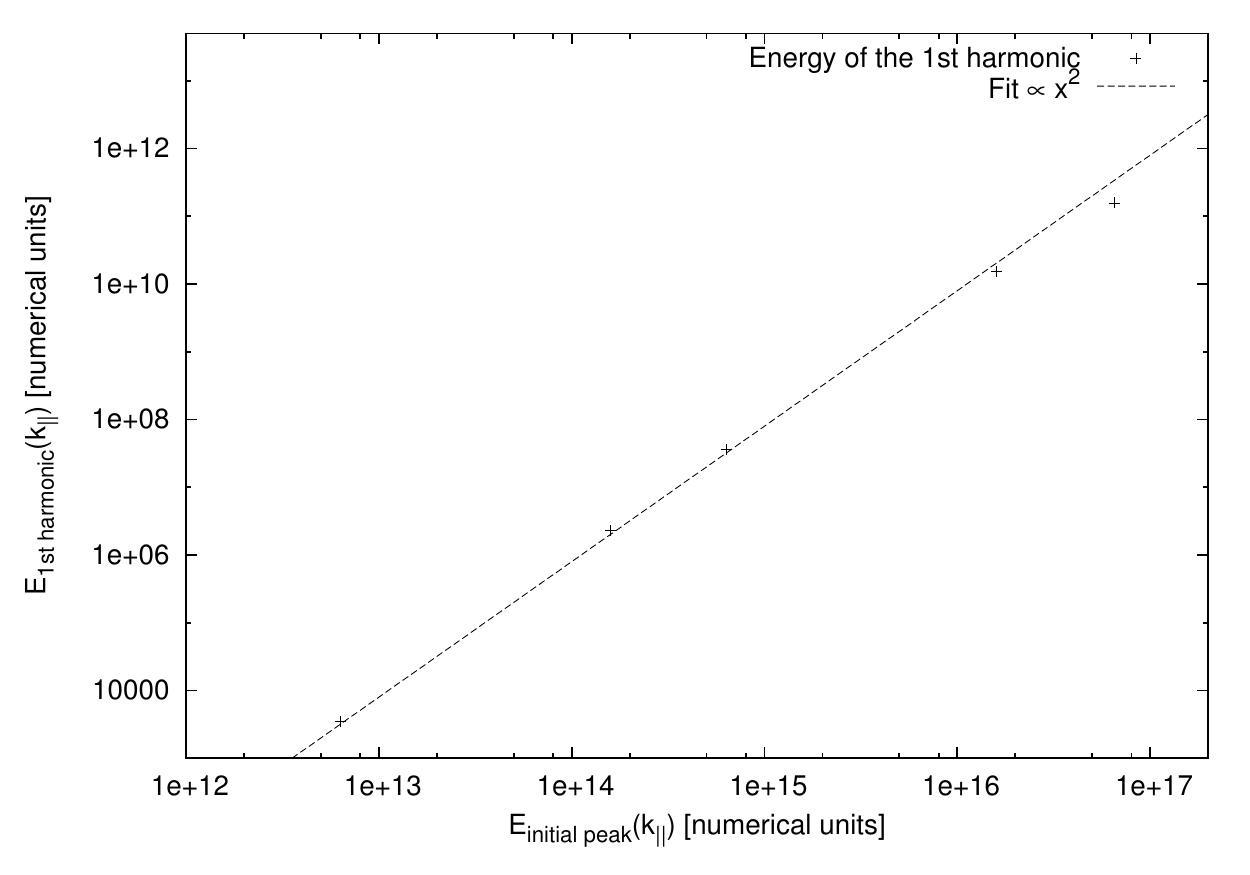} \caption{Energy dependency of the first harmonic on the initial peak energy. A fit resulted in a quadratic function. The simulation setup SI was used.}
\label{fig:peak-harmonics}
\end{center}
\end{figure}

During the evolution, higher harmonics of the initial peak arise. To investigate these in greater detail, we measured the energy of the initial peak and its first harmonic. The result is shown in Fig. \ref{fig:peak-harmonics}. Note that the generation of these modes starts at higher perpendicular wavenumbers (see Fig. \ref{fig:sphereplot-timeevo}, most picture left, first harmonic at $k_\perp \approx 15$) and not at purely parallel k.

In addition to the observed higher harmonics, another basic result of the simulation type SI is the discrete generation of other modes along the $k_\parallel$ axis especially at higher amplitudes. As seen in Fig. \ref{fig:peak24biggauss}, adjoining maxima develop next to the main Gaussian curve, e.g. at positions $k'_\parallel = 2 \pi \cdot 16$ and 38. The way the peak develops appears to vary with the amplitude, the peak with $\Gamma_1$ generates clearly fewer other modes than the larger peak with $\Gamma_2$. Both peaks show a significant change of thei original wavenumber position and a clear broadening. The two--dimensional spectra reveal a strong perpendicular development in k-space especially at large $|\vec{k}|$. During the decay the evolution becomes a little more isotropic, but the preferential direction of evolution is clearly perpendicular.

\begin{figure}[ht]
 \begin{center}
\includegraphics[width=\columnwidth]{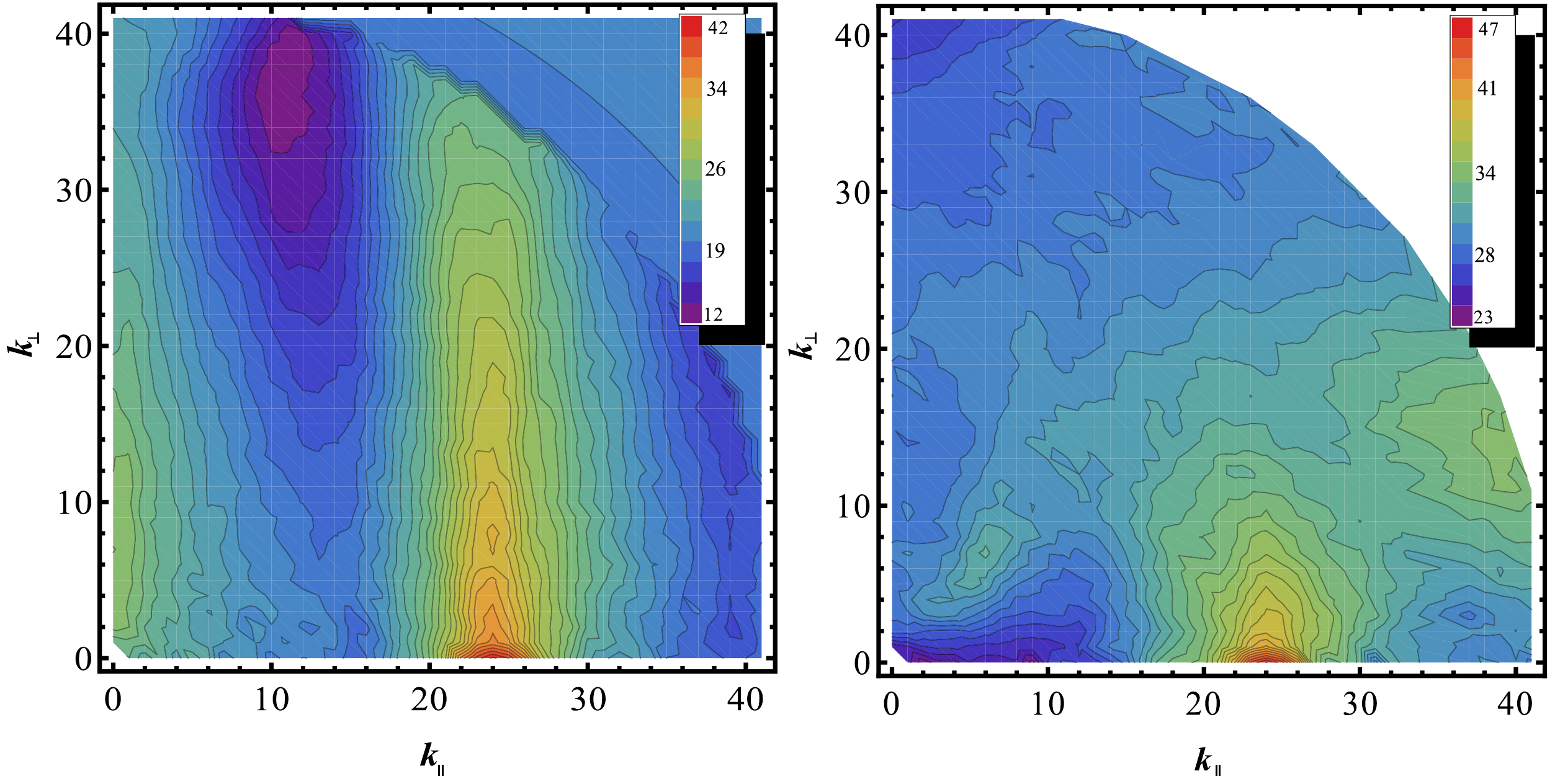} \caption{Simulation setup SI: Comparison between growth rates $\Gamma_1$ and $\Gamma_2$ at peak position $k'_\parallel = 2 \pi \cdot 24$ at maximum drive time. Although the effective energy input varies by a factor 100 (see table \ref{tab:peakamplitudes}) another transport mechanism becomes dominant. The smaller peak ($\Gamma_1$) develops dominantly in the perpendicular direction while the evolution of bigger peak ($\Gamma_2$) is more isotropic and tends towards smaller $k_\parallel$. The colours indicate the logarithm of the total spectral energy.}\label{fig:v31-24er-sphereplots}
\end{center}
\end{figure}

\begin{figure}[ht]
 \begin{center}
\includegraphics[width=\columnwidth]{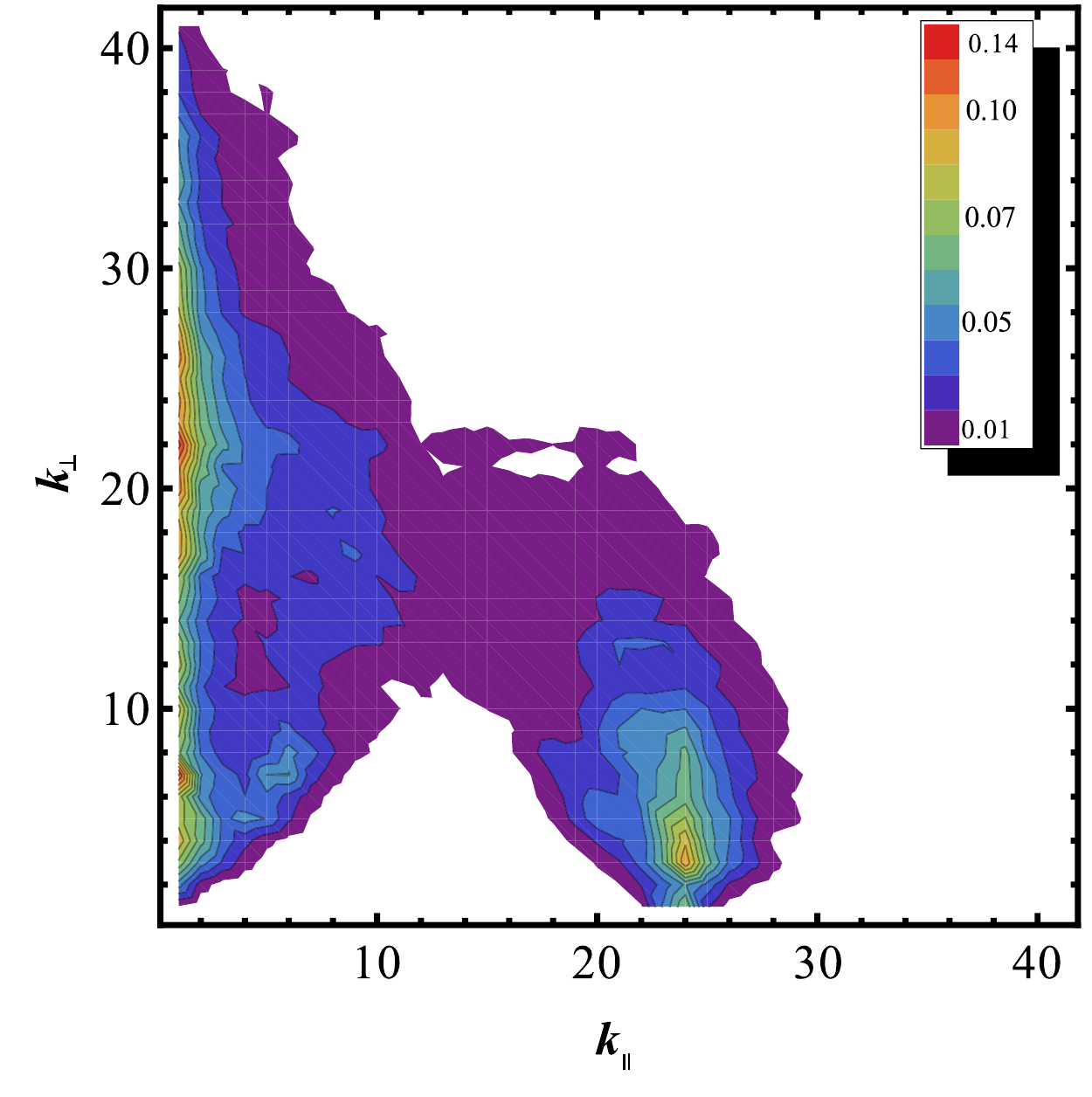} \caption{Map of the critical balance
parameter $\zeta$ for the right--hand plot in Fig. \ref{fig:v31-24er-sphereplots}. The contours linearly represent values between 0.01 and 0.14 of the integral values of $\zeta(k_\parallel,k_\perp)$. The peak structure and values near the $k_\perp$-axis are clearly visible.}\label{fig:v31-big24-critbalmap}
\end{center}
\end{figure}

The importance of the amplitudes $\Gamma_{1/2}$ is also visible in Fig. \ref{fig:v31-24er-sphereplots}. The development of each peak is very different. An interesting result is a more dominant  evolution of the $k'_\parallel = 2 \pi \cdot 24$ peak with the high growth rate $\Gamma_2$ that is towards smaller $k_\parallel$ directed. In the peak-dominated region the turbulence seems to increase. Therefore the $\zeta$ parameter (Eq. \ref{eq:zeta}) is of interest. Fig. \ref{fig:v31-big24-critbalmap} shows a map of values of $\zeta = [0.01 \cdots 0.15]$, i.e. near the critical balance. The same plot for the lower growth rate would be empty. This also indicates that the high $\zeta$ values along the $k_\perp$-axis in Fig. \ref{fig:v31-big24-critbalmap} stem from interactions with the peak.

The peaks at $k'_\parallel=2 \pi \cdot  8$ remain much longer than the higher modes, see table \ref{tab:peaktimes}. This is due to the higher $\vec{k}$ on which the dissipation process depends. Furthermore, the hyperdiffusivity damps higher modes more strongly ($\propto k^4$).

\begin{figure}[ht]
 \begin{center}
\includegraphics[width=\columnwidth]{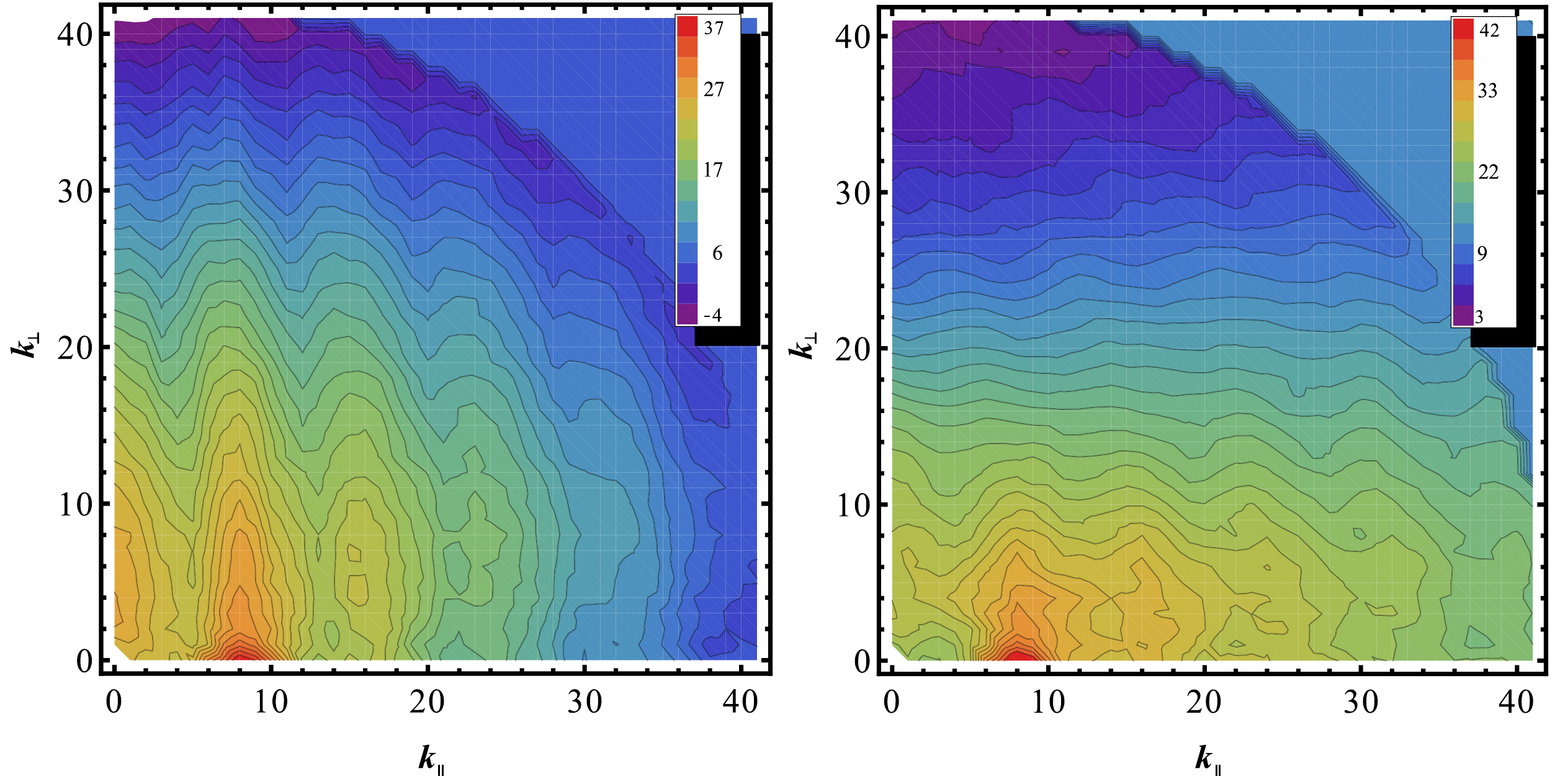} \caption{Two--dimensional peak evolution in setup SII at maximum drive time. The higher harmonics develop at lower $k_\perp$ compared to SI. The higher growth rate ($\Gamma_2$, right panel) shows a strong parallel evolution. The smaller amplitude ($\Gamma_1$, left panel) develops dominantly towards higher $k_\perp$. The colours indicate the logarithm of the total spectral energy.}\label{fig:v34-8er-sphereplots}
\end{center}
\end{figure}

\begin{figure}[ht]
 \begin{center}
\includegraphics[width=\columnwidth]{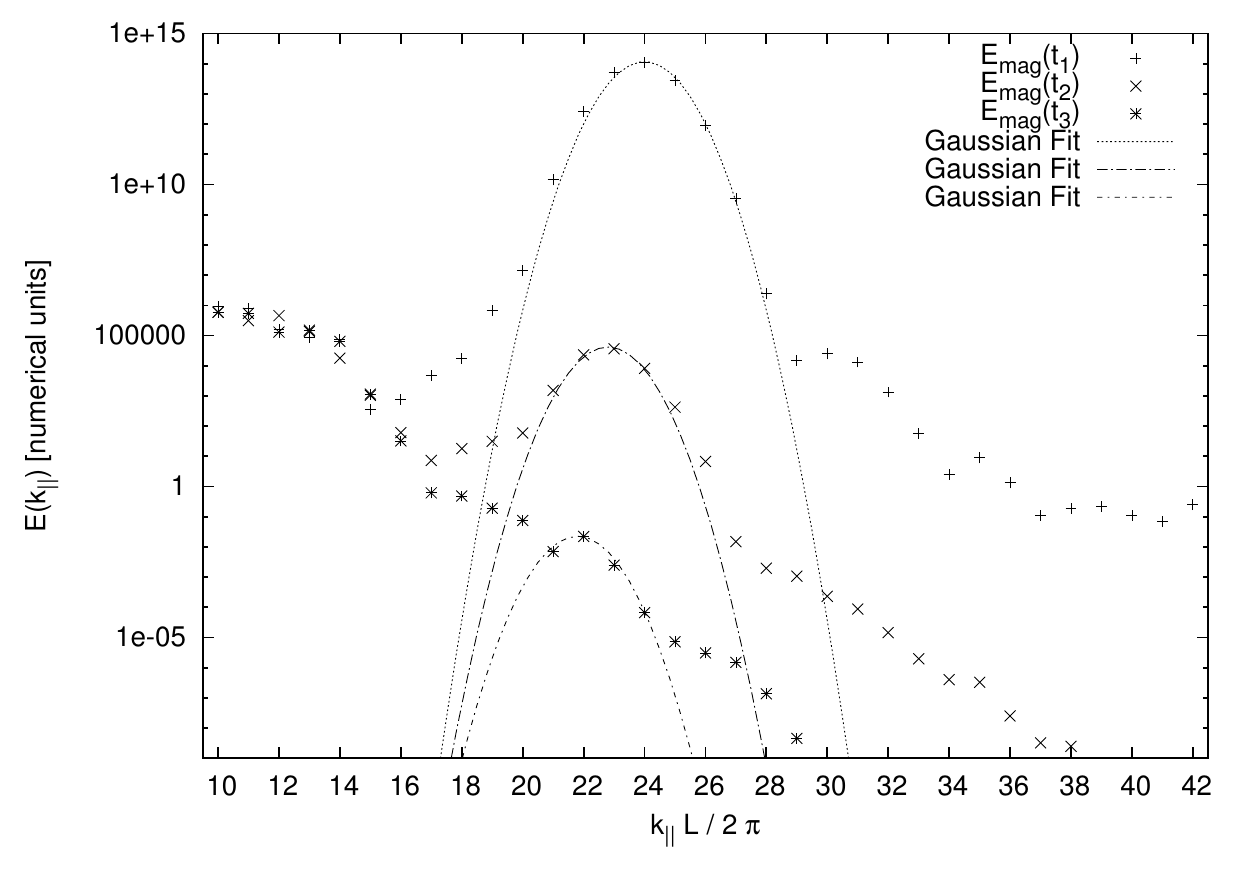} \caption{Simulation setup SII: Time evolution of a Gaussian distributed amplification at $k'_\parallel=2\pi \cdot 24$ at the smaller growth rate $\Gamma_1$. Again a 1D cut spectrum along $k_\parallel$ axis is shown. The shift of the peak is stronger compared to SI.}\label{fig:v34-peak24smallgauss}
\end{center}
\end{figure}

The simulation of the same peaks in setup SII with higher $\nu$ reveals one key feature. The shift towards smaller $k_\parallel$ positions occurs on both, $k'_\parallel=2\pi \cdot 8$ and 24 peaks and is stronger compared to SI. The shifting is significant, especially at $k'_\parallel=2\pi \cdot 24$, which is shown in Fig \ref{fig:v34-peak24smallgauss}. Within 2.55 s the original position changes from $k'_\parallel=2\pi \cdot 24$ to 22. The peak amplitude is again important for the evolution. We observed that higher growth rates lead to an isotropic evolution, whereas the lower growth rates show strongly perpendicular development. The effective peak energy is lower and consequently the energy transport towards higher wavenumbers is more restricted. There are also fewer of higher harmonics. This can be observed by direct comparison of the left plot in Fig. \ref{fig:v34-8er-sphereplots} with the middle plot in Fig. \ref{fig:sphereplot-timeevo}. As expected, the decay of the energy is faster compared 
to SI (see table \ref{tab:peaktimes}).

\begin{figure}[ht]
 \begin{center}
\includegraphics[width=\columnwidth]{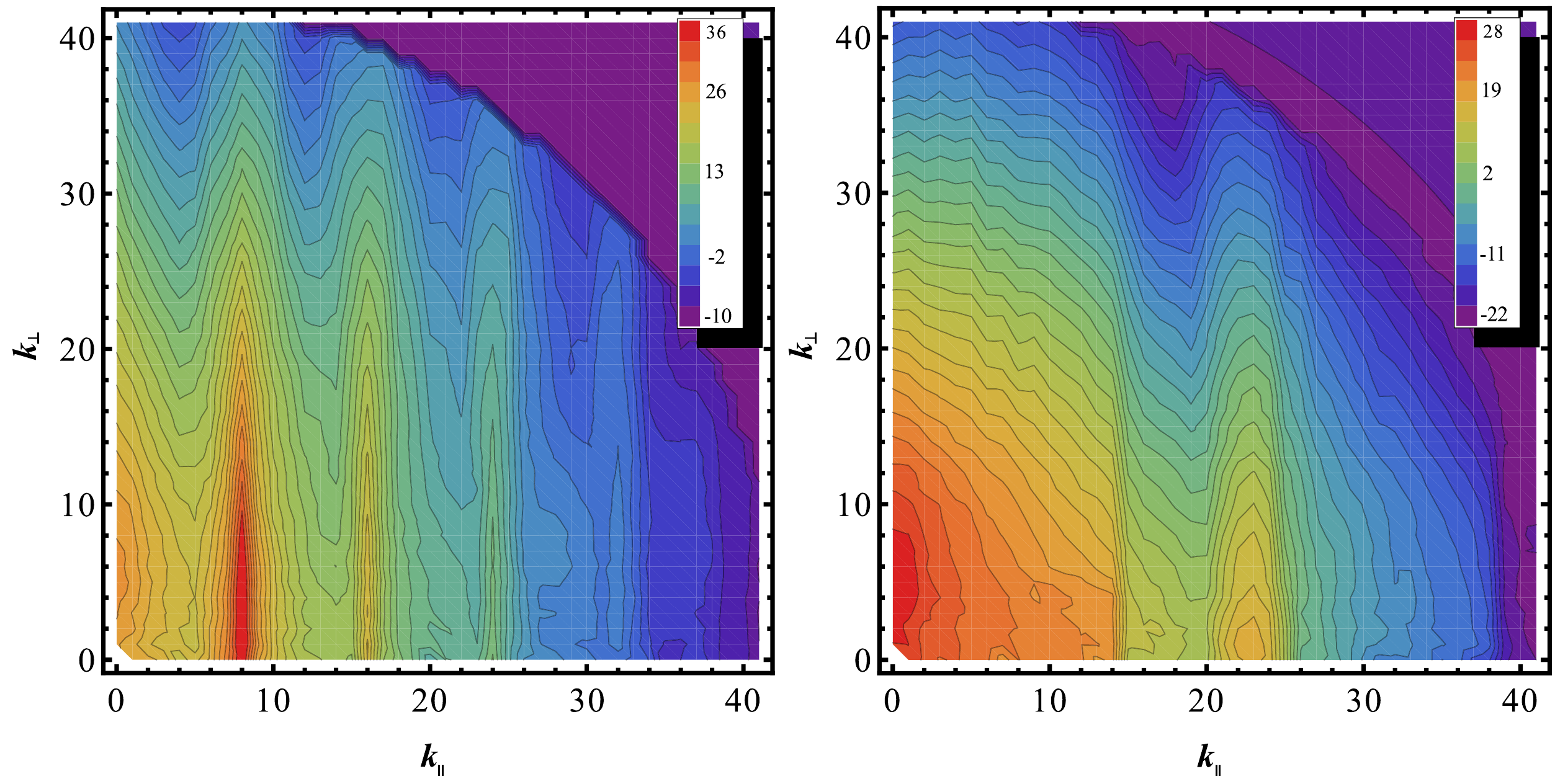} \caption{Evolution of the two peak positions at $t_\text{mid}$ in setup SIII. Very strong perpendicular development in all SIII simulations is observed. The edge of the parallel driving range at $k'_\parallel=2\pi \cdot 14$ is clearly visible in the right--hand panel.}\label{fig:v35-8u24-sphereplots}
\end{center}
\end{figure}

The simulation setup SIII reveals very strong perpendicular evolution for all peaks. Two examples are given in Fig. \ref{fig:v35-8u24-sphereplots}. The growth rates seem not to have a strong influence on the development. Only more higher harmonics of $k'_\parallel=2\pi \cdot 8$ peak are visible with $\Gamma_2$. The time of total energy decay of the peaks is slightly longer than in the simulations SII.

\begin{figure}[ht]
 \begin{center}
\includegraphics[width=\columnwidth]{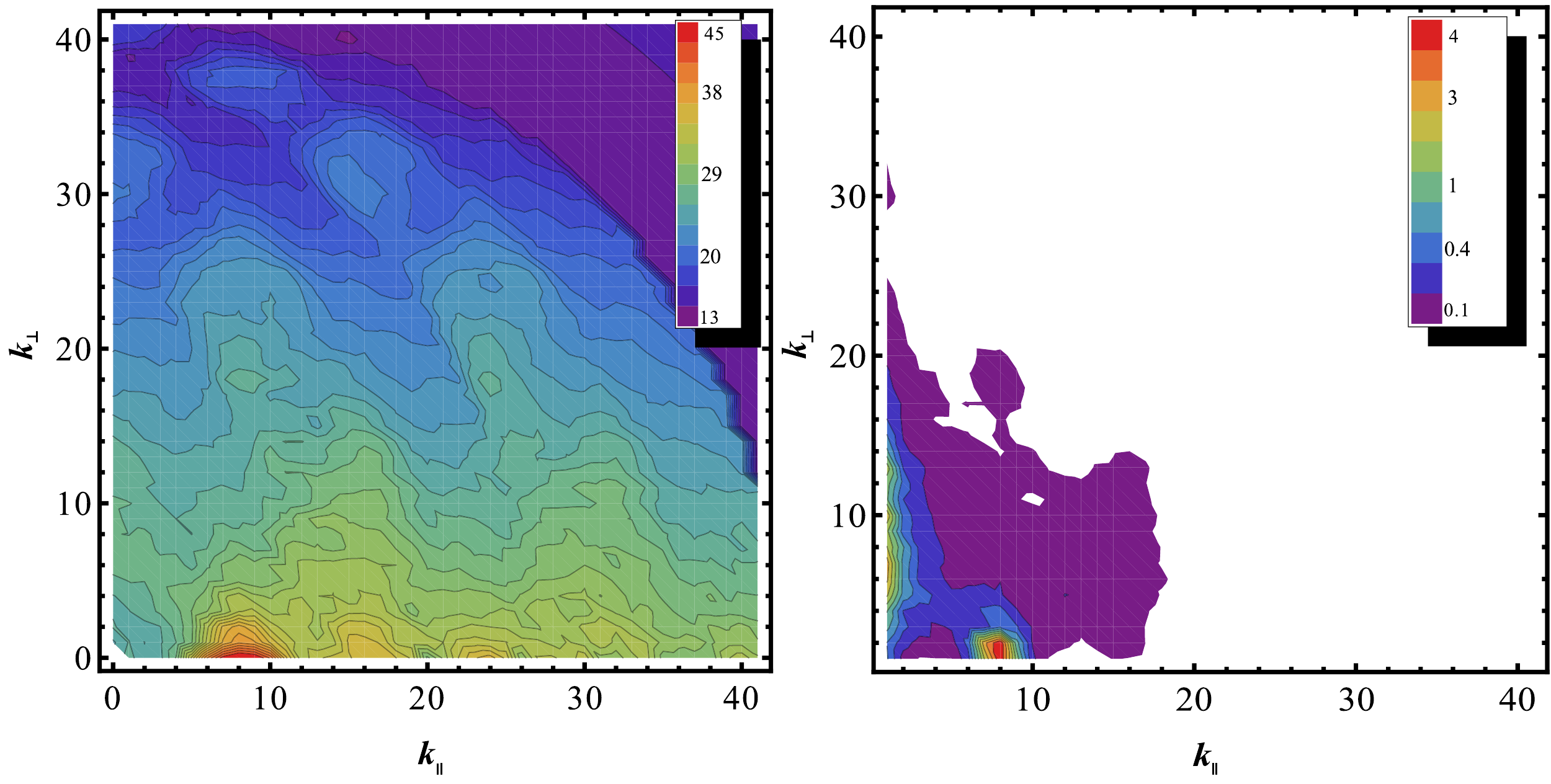} \caption{Two--dimensional peak evolution in setup SIV. Left: Development of the peak at $k'_\parallel=2\pi \cdot 8$ with growth rate $\Gamma_2$. The figure presents the state 1.28 s after the maximum driven peak ($t_0$). Right panel: Corresponding map of the critical balance parameter $\zeta$. Each contour represents integral values above $\zeta=0.1$. The colour scaling is linear.}\label{fig:v36-big8_u_critbal}
\end{center}
\end{figure}

In the last simulation setup SIV the magnetic background field was reduced by two orders of magnitude. The resulting $dB/B_0$ ratio is of the order of 10 and the turbulence development is highly isotropic. The peaks at $k'_\parallel=2\pi \cdot 8$ and $k'_\parallel=2\pi \cdot 24$ both at growth rate $\Gamma_2$ show very interesting features.

\begin{figure}[ht]
 \begin{center}
\includegraphics[width=\columnwidth]{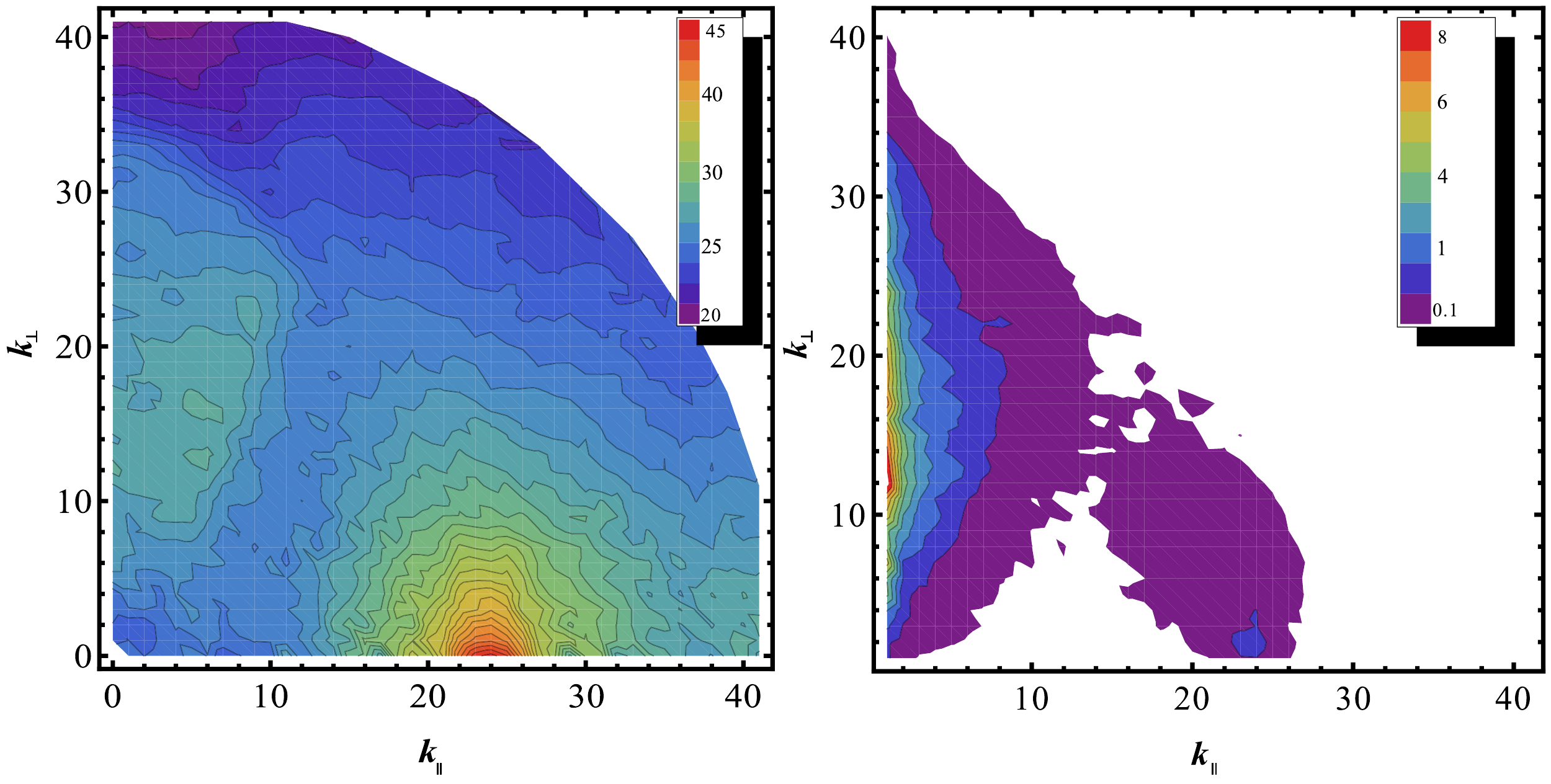} \caption{Two--dimensional peak evolution in setup SIV. Left: Development of the peak at $k'_\parallel=2\pi \cdot 24$ with growth rate $\Gamma_1$. The figure presents the state 0.68 s after the maximum driven peak ($t_0$). Right panel: Corresponding map of the critical balance parameter $\zeta$. Each contour represents integral values above $\zeta=0.1$.The colour scaling is linear.}\label{fig:v36-small24_u_critbal}
\end{center}
\end{figure}

In addition to the typical peak evolution and generation of higher harmonics, other structures also arise at high $k_\perp$. As shown in the Figs. \ref{fig:v36-big8_u_critbal} and \ref{fig:v36-small24_u_critbal}, both peak positions generate these structures, but at various positions. During the development of the peak at position $k'_\parallel=2\pi \cdot 8$ theses structures arise perdominantly at high $(k'_\parallel,k'_\perp)$ locations, e.g. $(2\pi \cdot 10,2\pi \cdot 18)$, $(2\pi \cdot 10,2\pi \cdot 38)$, $(2\pi \cdot 18,2\pi \cdot 32)$ and $(2\pi \cdot 24,2\pi \cdot 18)$, see Fig \ref{fig:v36-big8_u_critbal}. The structures evolving with the $k'_\parallel=2\pi \cdot 24$ simulations are instead located at middle $k_\perp$ but low $k_\parallel$, e.g. $(2\pi \cdot 5,2\pi \cdot 13)$, $(2\pi \cdot 5,2\pi \cdot 17)$ and $(2\pi \cdot 8,2\pi \cdot 22)$, see Fig \ref{fig:v36-small24_u_critbal}. In contrast to the higher harmonics, these structures are not necessarily integral multiples of the initial peaked mode. A 
map of the critical balance parameter $\zeta$ was calculated for both plots. Around the wavenumber of the peaks and along the $k_\perp$ axis this parameter is of order 1. Interestingly, during the $k'_\parallel=2\pi \cdot 24$ simulation this parameter increases, especially at the $k_\perp$ axis.
In contrast to the other setups, the higher harmonics are located along the $k_\parallel$ axis at low or zero $k_\perp$. A significant shift along this axis towards smaller parallel wavenumbers is observed, e.g. as shown in Fig. \ref{fig:vglsmallpeak8setups}.

We discuss possible explanations for these phenomena below.

\section{Discussion}\label{sec:discuss}

As discussed in Sect. \ref{sec:theory}, there are three possibilities how an excited wavemode can develop: through diffusion, dissipation and convection.

A general conclusion from our simulations is that the dynamics of these mechanisms are strong at high wavenumbers. Especially for the dissipation process this is not unexpected because of it strongly depends on th wavenumber. The hyperdiffusivity might amplify this effect because of the higher power in $k$. The Figs. \ref{fig:peak24smallgauss} and \ref{fig:peak24biggauss} clearly show a rapid dissipation of energy because the peak loses amplitude very fast. Also table \ref{tab:peaktimes} indicates a decay in short timescales at high wavenumbers. However, a broadening also arises at the flanks of the Gaussian distribution. The broadening is strong for the higher growth rate $\Gamma_2$. Within a time interval of 1.7 s after the driving range the FWHM is increased by roughly 70\% at $\Gamma_2$, whereas the peak with smaller amplitude is broadened by ca. 15 \%. A possible explanation is the equilibrium between energy- and enstrophy cascade, which causes a similar flow of energy to large and small wavenumbers \citepads{mininni09}.

\begin{figure}[ht]
 \begin{center}
\includegraphics[width=\columnwidth]{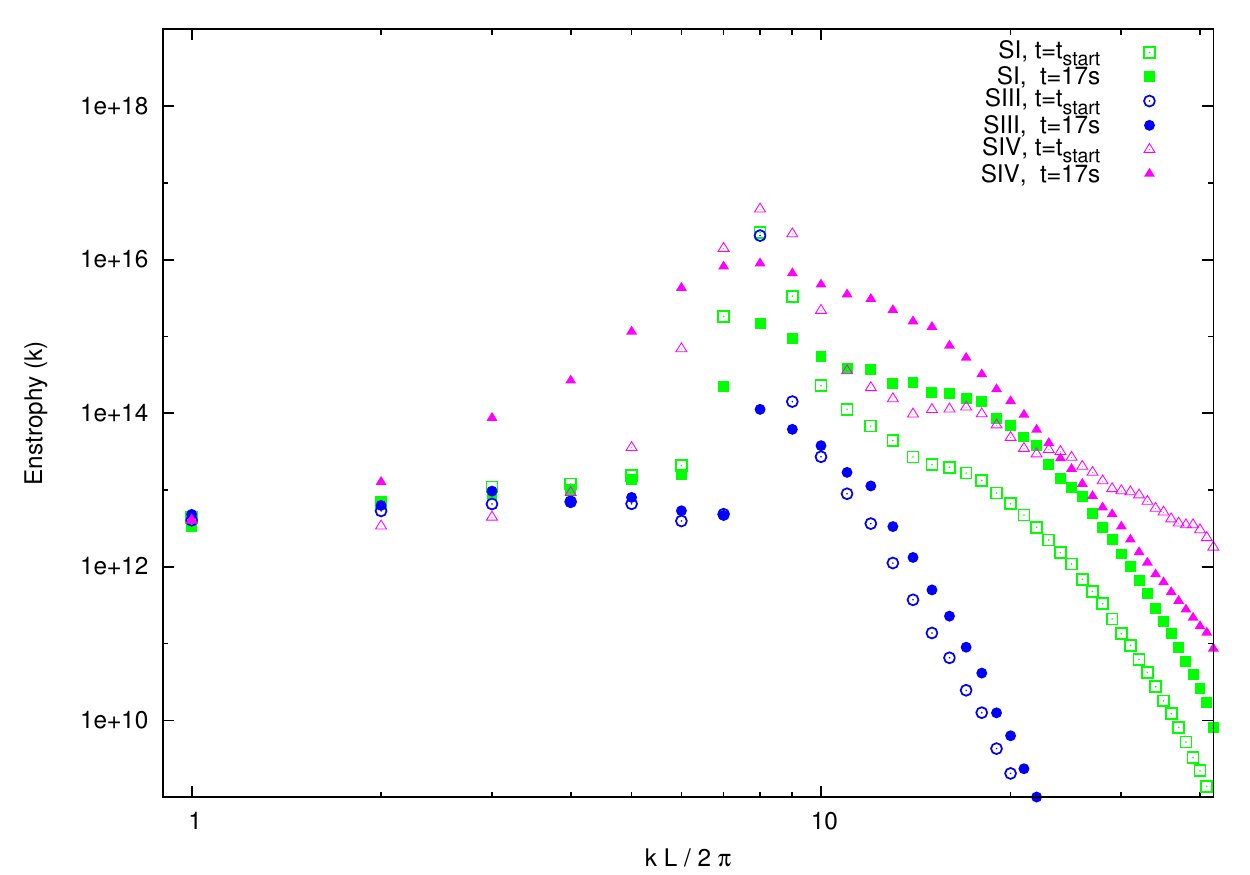} \caption{Time evolution of the enstrophy for the simulation setups SI, SIII and SIV. The strong magnetic background field of SIII prevents the enstrophy from developing while a clear change in SIV is visible.}\label{fig:enstrophy}
\end{center}
\end{figure}

The convective mechanism shifts the maximum of the peak. Convection is a slow process compared to dissipation and diffusion within the simulated regime. Nevertheless we were able to observe it within our simulations, e.g. in Fig. \ref{fig:peak24smallgauss}. The transport is towards smaller wavenumbers, which indicates an enstrophy cascade. This effect is more typical for two--dimensional plasmas. The MHD-development in anisotropic plasmas is mostly effectively two--dimensional. This lead to inverse energy cascades as well as upscaling enstrophy cascades. These mechanisms generate larger vortices and consequently transfer energy to smaller wavenumbers.

\begin{figure}[ht]
 \begin{center}
\includegraphics[width=\columnwidth]{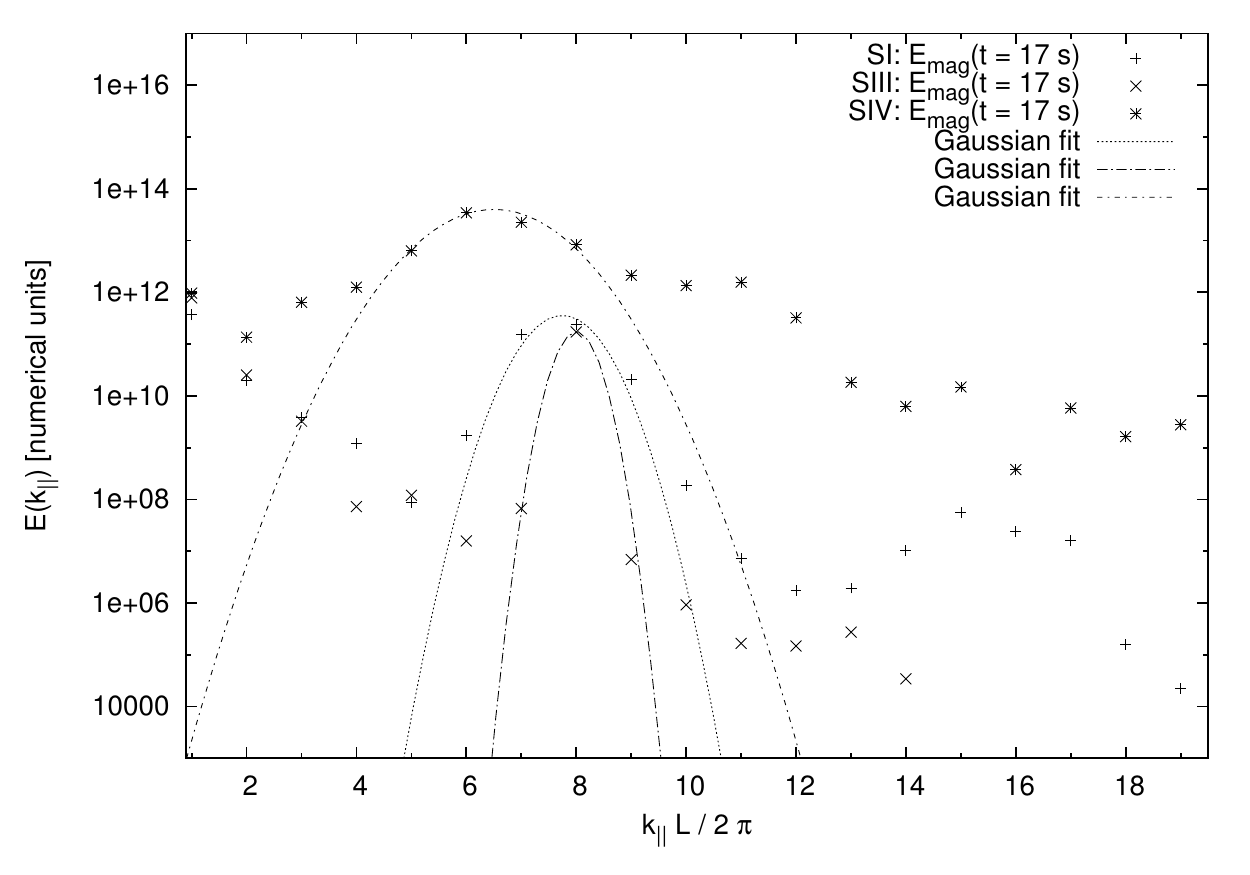} \caption{Comparison of the convective peak shifting between the setups with changing background field $B_0$ SI, SIII and SIV. The strong $B_0$ in SIII causes no observable shifting at all, whereas the weak $B_0$ leads to a significant change of the original peak position. One possible explanation is an enstrophy cascade.} \label{fig:vglsmallpeak8setups}
\end{center}
\end{figure}

To investigate this behaviour in more detail, we concentrated on the peak $k'_\parallel = 2 \pi \cdot 8$ with growth rate $\Gamma_1$ within the setups of changing magnetic background field SI, SIII and SIV. The comparsion of these setups  is shown in Fig. \ref{fig:vglsmallpeak8setups}. We observe a slight shift of the in setup SI after 17s of the maximum drive from position 8 to 7.7. The simulation SIII with increased $B_0$ shows no transport of the peak. After the same time interval it is still at position 8. The evolution in setup SIV with a small $B_0$ is very strong. After 17s the peak has shifted by roughly 1.5 grid positions from $k'_\parallel = 2 \pi \cdot 8$ to 6.5. The development of an enstrophy cascade explains this behaviour. The strong magnetic field effectively makes the MHD-evolution  one-dimensional. As $B_0$ decreases, the evolution becomes less restricted to the magnetic background field. Consequently, the enstrophy cascade increases. Fig. \ref{fig:enstrophy} clearly supports this 
explanation. The enstrophy was calculated by
\begin{align}
 \varepsilon(\vec k) = \int \text{d}^3 k \; |\vec k\times \vec u|^2.
\end{align}

All two--dimensional spectra show a strong evolution in the perpendicular direction. This is consistent with the theories of  \citetads{gsweak} and \citetads{gsstrong} within a turbulent plasma, as explained in Sect. \ref{sec:theory}.
The process of this perpendicular evolution is caused by the energy cascade. Mainly Fig. \ref{fig:v35-8u24-sphereplots} shows strong perpendicular behaviour, whereas the evolution clearly becomes more isotropic in SIV (see Figs. \ref{fig:v36-big8_u_critbal} and \ref{fig:v36-small24_u_critbal}). This is due to the increasing strength of the turbulence for the cascade, which is expressed by the $dB/B_0$ ratio.

The dissipation coefficients presented in table \ref{tab:diffcoeffs} do remain constant especially for all simulations with $\Gamma_2$ at $k'_\parallel = 2 \pi \cdot 8$. This implies that spatial diffusion is the dominant process for these simulations. The most significant change of the dissipation coefficient is at $k'_\parallel = 2 \pi \cdot 24$ for setup SII, between $\tau_1$ and $\tau_2$. We interpret this as a nonlinear effect where wavemodes are generated, which in turn triggers the cascade. This influences the energy of the initialised Gaussian very strongly.
As expected, the dissipation coefficients for setup SII are stronger for the $k'_\parallel = 2 \pi \cdot 24$ peaks. The spatial diffusion coefficient is connected with the wavenumber dissipation process via $\nu$. This is because the spatial diffusion process is the only mechanism that leads to energy losses in the k space (see Sect. \ref{sec:theory}). This is at least valid for wavemodes far below the antialiasing edge, where energy is artificially removed as well. A connection of wavenumber diffusion to $\nu$ is observed in Fig. \ref{fig:v34-8er-sphereplots} where the Gaussian shape is significantly broadened, which is caused by the diffusion process.

The observed higher harmonics could resemble a three-wave process $k_{8}+k_{8} \rightarrow k_{16}$. This is supported by Fig. \ref{fig:peak-harmonics}. The energy dependency between peak and the next higher harmonic is quadratic. This is also the case for three-wave interactions \citepads{1969npt..book.....S}. As pointed out before, this is forbidden for Alfv\'en waves by wave interaction processes. Just oppositely directed wavepackages can collide, hence momentum conservation would be violated by the proccess described above. This wave interaction can only take place with oppositely directed waves of the background plasma. This must also be true since a cross-check simulation of the peak without turbulent background did not show these harmonics. Another explanation is given by \citetads{galinsky97}: Alfv\'en waves interact with themselves, which leads to wave steepening.

Investigating the strong turbulence evolution in simulation SIV shows unexpected structures at high $k_\perp$ (see Fig. \ref{fig:v36-big8_u_critbal} and \ref{fig:v36-small24_u_critbal}).
This effect might be caused by the critical balance of strong turbulence within the Goldreich and Sridhar description. This requires the parameter
$\zeta$ to be $\sim 1 $, which is the case for locations in the vicinity of the peaks and along the perpendicular axis. Mainly the $k'_\parallel=2\pi \cdot 24$ peak seems to amplify the region at $k'_\perp =2\pi \cdot 12$ and $k'_\parallel = 0$. Values of $\zeta > 0.1$ lead to the development of these structures, but it is not possible to conclude whether the stuctures arise first and then generate higher values of $\zeta$, or vice versa. Nevertheless, the turbulence strength increases within these regions, which agrees with \citetads{gsstrong} where $\zeta$ is assumed to become unity during the turbulence evolution. In addition to setup SIV, SI also shows this behaviour as presented in Fig. \ref{fig:v31-24er-sphereplots} and \ref{fig:v31-big24-critbalmap}. More investigation is needed to clarify the underlying processes.

\section{Conclusion}

We analysed the evolution of waves generated by proton beams in a turbulent medium. This evolution may play an important role in diffusive shock acceleration in the heliosphere. Our study has revealed that three different processes as sketched in Fig. \ref{fig:peak-evo-sketch} are taking place.

The most interesting question is wheter wavemodes excited by particles of a certain energy yield wavemodes that interact with other particle energies. The observed shifting of the initial parallel wavenumber position towards smaller $k_\parallel$ influences the particle acceleration at these modes. As shown in Eq. \ref{eq:wave-particle-res} and in the corresponding section, this means that higher energetic particles can be accelerated because the wavemode develops towards higher spatial scales \citepads{2007ApJ...658..622V}. The shift towards smaller $k_\parallel$ is indeed fairly minor in this simulation. This is because of the injection of energy at only one single wavenumber and the limited simulation time. We also expect an effect through nonlinear amplification: Waves at smaller $k_\parallel$ accelerate higher energetic particles, again injecting energy at lower $k_\parallel$. On the other hand a strong evolution towards high $k_\parallel$ has also been observed in terms of development of 
higher harmonics of the initialised mode. Consequently, particle acceleration at lower energies is also possible. It should be noted, however, that this is not consistent with isotropic diffusion of energy in wavenumber space as assumed in \citetads{2007ApJ...658..622V}. This means that the previous models of the wave-particle system will have to be updated accordingly to account for the strong dependence of energy transport on the direction in wavenumber space. Especially the development of the higher harmonics contradicts an identical forward and backward cascade.

The strong perpendicular evolution of the  peak initialised with purely parallel propagating waves causes higher orders ($|m|>1$, see Eq. \ref{eq:wave-particle-res}) of resonance between solar particles and the amplified mode. This is because Eq. \ref{eq:growthraterami} has to be modified in this case \citepads{Schlickeiser2002}.

Owing to limited computational power we have not been able to investigate the effect of critical balance in detail. But we note that under certain conditions ($k_\parallel/k_\perp$, amplitude) the evolution may be governed by the critical balance.

\begin{acknowledgements}
We express our graditude to Rami Vanio, Timo Laitinen and Markus Battarbee for cooperation and their contributions to this work.\\
We acknowledge support from the Deutsche Forschungsgemeinschaft through grant SP 1124/3.\\
SL additionally acknowledges support from the European Framework Program 7 Grant Agreement SEPServer - 262773.\\
We thank the anonymous referee for her/his detailed comments, which improved the paper significantly.
\end{acknowledgements}

\bibliographystyle{aa}
\bibliography{ref}

\end{document}